\documentclass{article}

\usepackage{arxiv}

\usepackage[utf8]{inputenc} 
\usepackage[T1]{fontenc}    
\usepackage{hyperref}       
\usepackage{url}            
\usepackage{booktabs}       
\usepackage{amsfonts}       
\usepackage{nicefrac}       
\usepackage{microtype}      
\usepackage{lipsum}

\usepackage{amsmath}
\usepackage{multirow}
\usepackage{graphicx}
\usepackage{subcaption}
\usepackage{authblk}
\usepackage[%
    style=numeric-comp,sorting=none,
    sortcites=true,doi=false,url=false,
    giveninits=true,hyperref]{biblatex}
\makeatletter
\g@addto@macro{\UrlBreaks}{\UrlOrds}
\makeatother

\DeclareUnicodeCharacter{2215}{/}

\title{Learning to Approximate Functions Using Nb-doped SrTiO$_3$ Memristors}

\author[1,2]{ Thomas F.~Tiotto }
\author[1,3]{ Anouk S.~Goossens }
\author[1,2]{ J.~P.~Borst }
\author[1,3]{ T.~Banerjee }
\author[1,2]{ N.~A.~Taatgen }
\affil[1]{Groningen Cognitive Systems and Materials Center, University of Groningen, Groningen, The Netherlands}
\affil[2]{Bernoulli Institute for Artificial Intelligence, University of Groningen, Groningen, The Netherlands}
\affil[3]{Zernike Institute for Advanced Materials, University of Groningen, Groningen, The Netherlands}
\affil[ ]{\texttt{\{t.f.tiotto,a.s.goossens,j.p.borst,t.banerjee,n.a.taatgen\}@rug.nl}}

\begin{document}
\maketitle

\begin{abstract}
Memristors have attracted interest as neuromorphic computation elements because they show promise in enabling efficient hardware implementations of artificial neurons and synapses. 
We performed measurements on interface-type memristors to validate their use in neuromorphic hardware.
Specifically, we utilised Nb-doped SrTiO$_3$ memristors as synapses in a simulated neural network by arranging them into differential synaptic pairs, with the weight of the connection given by the difference in normalised conductance values between the two paired memristors.
This network learned to represent functions through a training process based on a novel supervised learning algorithm, during which discrete voltage pulses were applied to one of the two memristors in each pair.
To simulate the fact that both the initial state of the physical memristive devices and the impact of each voltage pulse are unknown we injected noise at each timestep. 
Nevertheless, discrete updates based on local knowledge were shown to result in robust learning performance.
Using this class of memristive devices as the synaptic weight element in a spiking neural network yields, to our knowledge, one of the first models of this kind, capable of learning to be a universal function approximator, and strongly suggests the suitability of these memristors for usage in future computing platforms.
\end{abstract}

\keywords{neuromorphic computing \and supervised learning \and interface memristor \and Nb-doped SrTiO$_3$ \and neural networks \and spiking neural network \and function approximation}

\section{Introduction}
The field of Machine Learning is, at its core, concerned with building function approximators from incomplete data samples. 
The state of the art approach to solving this problem is using artificial neural networks (ANNs), where a large number of real-valued artificial neurons are connected to each other by means of weights.
The neurons in such networks are typically arranged into multiple layers, and are therefore referred to as \textit{deep learning}.
The optimisation process is performed by updating the weight matrices defining the connection weights between pairs of neurons and is guided by learning rules, which are heuristic optimisation algorithms capable of iteratively tuning the network weights to minimise some error function. 
This process is based on either global (as in the classic back-propagation algorithm) or local knowledge (which is more biologically plausible); the typical outcome is an interpolation for the hidden mapping from input samples to observed outputs.

Traditional neural networks usually require considerable power and are not necessarily constrained by physical limitations. 
One of the main reasons for this shortcoming is that artificial neural networks are run on traditional Von Neumann architectures in which memory and computation are not \textit{co-located}.
The high energy required by deep learning can be ascribed to the fact that an artificial neural network is essentially a non-Von Neumann computational model, where memory and computation are co-located in connection weight matrices, being simulated on hardware that implements a different computational paradigm.

Even though many important advances in deep learning have been “biologically inspired” (e.g., convolutional neural networks), it is unclear how far the current deterministic approach can progress because of energy requirements, architectural complexity, and capacity to generalise.

To reduce energy, alternative approaches to traditional neural networks aim to implement learning in neuromorphic hardware. 
Memristors are novel fundamental circuit elements that have attracted a great deal of interest for this purpose because they exhibit multilevel conductance states, giving them - for example when arranged in a crossbar array architecture - the ability to carry out parallel vector-matrix multiplication ( $C = \vec{a} B$ ), which is the most important mathematical operation underpinning artificial neural network implementations (\cite{10.1038/s41563-019-0291-x}).
The ability to carry out this operation, where the input is voltage $V$, output is current $I$, and the weight corresponds to conductance $G$ (the inverse of resistance $R$), follows from Ohm's law $I=VG$ and Kirchoff's current law $I_j=\sum I_i$.
Due to their non-volatile nature and ability to co-locate memory and computing, memristors can enable parallel physical computation within a low energy envelope.

Memristors have also been used as direct hardware implementations of weights in order to replicate standard deep learning software architectures, such as long short-term memory (LSTM) (\cite{10.1038/s42256-018-0001-4}).
Such systems can then be trained using standard gradient-based machine learning algorithms but in doing so limit themselves to the mainstream computational paradigm and do not take full advantage of the material properties of the memristive devices.

Another alternative approach to computation in which the use of memristors is currently being explored is that of \textit{reservoir computing} (\cite{10.1016/j.cosrev.2009.03.005}).
A reservoir is a collection of recurrently interconnected units exhibiting short-term memory and non-linear response to inputs; the connections of the reservoir are fixed and the training of the network is focused on a simple feed-forward readout layer thus reducing training time and complexity. 
Previous literature has shown that memristor-based reservoir computers are capable of excellent performance on a variety of tasks by taking advantage of the intrinsic characteristics of the physical devices.
A recurrent connection is a way of adding temporal dynamics to a model, but a memristor already is such a system.
Existing memristor-based reservoirs have thus eschewed the recurrent connections between the reservoir units and relied on the fact that programming pulses at different moments in time have varying effect on the state of the device (\cite{10.1038/s41467-017-02337-y,moon2019temporal,aisy.201900084}).
\textit{Diffusive} memristors - whose memristance is governed by fast diffusive species - have also seen use as reservoir units (\cite{wang2017memristors}), but also as artificial neurons (\cite{10.1038/s41928-018-0023-2}), as their dynamics are well-suited to representing the leaky integrate-and-fire model of neuronal computation.

The biological parallel with the brain can be taken further by substituting continuous idealised neurons, found in traditional artificial neural networks, for their spiking equivalent and by using a learning algorithm which updates weights on the basis of local knowledge.
For example, considerable interest has been devoted to using memristors to implement Hebbian learning rules which are thought to underpin the capacity of the brain to learn in an unsupervised manner and adapt to novel stimuli.
Multiple flavours of spike timing-dependent plasticity (STDP) and hardware architectures have been proposed as ways to make good use of the physical characteristics of memristors, by appropriately shaping voltage pulses (\cite{10.3389/fnins.2013.00002,ambrogio2016}), or by
controlling second-order state variables (\cite{10.1021/acs.nanolett.5b00697}).


The most widely studied memristors for neuromorphic applications are ones relying on electric field control of nanoscale filaments between two electrodes (\cite{seo2011,wang2018,moon2018,hu2018}), phase transitions (\cite{kuzum2012,ambrogio2016,wang2017}) or voltage induced control of the ferroelectric configuration (\cite{nishitani2012,oh2019,kim2019}). 
A less studied class of materials for synaptic device applications are interface-type memristors where electric field controlled resistive switching results from changes occurring at interfaces. 
While the resistive switching in for example metal/Nb-doped SrTiO$_3$ (Nb:STO) Schottky junctions, is well documented in literature, there are not many reports in which they are considered as individual neuromorphic components (\cite{yin2016,jang2018,zhao2019}) and, to the best of our knowledge, no reports of their use as components in neural networks. 
Often, memristive devices require forming processes, which can be unfavourable for device performance and network integration (\cite{amer2017,Kim2019EF}).
An attractive feature of Nb:STO memristors is that the switching behaviour is present in the as-fabricated device. 
In addition, they also show reproducible multi-level resistive switching and low reading currents at room temperature, making them potentially ideal as components in neuromorphic computers.
Such computers need to be able to deal and operate within the physical constraints of the particular class of memristor they utilise as hardware substrate.
The heterogeneous characteristics of memristive devices also entails that a neuromorphic computer be aware of their physical characteristics and responses, in order to be able to utilise them to their full potential.

Here, we utilised Nb:STO memristors as synapses in a simulated spiking neural network to explore the suitability of this class of memristors as computing substrate. 
Measurements were conducted on physical devices, in which the change in resistance in response to a series of voltage pulses was monitored. 
We found the resistance values in response to forward voltage pulses to follow a power law, whereby each pulse gives rise to an amplitude-dependent decrease in the resistance.
Next, we built a simulated model using the Nengo Brain Builder framework, consisting of pre- and post-synaptic neuronal ensembles arranged in a fully-connected topology.
Each artificial synapse was composed of a \textit{differential synaptic pair} of memristors, one designated as “positive” and one as “negative”, and the weight of the connection was given by the difference in normalised conductance values between the two paired memristors.
Using a training process during which discrete voltage pulses were applied to one of the two memristors in each pair, we showed that our model was capable of learning to approximate transformations from randomly and periodically time-varying signals, suggesting that it could operate as a general function approximator.
The initial state of the memristive devices was unknown, as were their exact operating parameters due to the addition of noise; this simulated the same kind of constraints faced in hardware, making it likely that the learning model could be ported to a physical implementation.
Most importantly, robust learning performance was proven using only discrete updates based on local information.


\section{Materials and Methods}

\subsection{Device Fabrication}

Memristive devices are material systems which undergo reversible physical changes under the influence of a sufficiently high voltage, giving rise to a change in resistance (\cite{williams2008,spiga2018}).
Here, metal/Nb-doped SrTiO$_3$ devices were used, where resistive switching results from changes occurring at the Ni/Nb:STO interface (\cite{sim2005,seong2007,sawa2008,rodenbucher2013,mikheev2014,goossens2018}).
We fabricated devices on (001) SrTiO$_3$ single crystal substrates doped with 0.01 wt\% Nb in place of Ti (obtained from Crystec Germany). 
To obtain a TiO$_2$ terminated surface, a wet etching protocol was carried out using buffered hydrofluoric acid, this was followed by in-situ cleaning in oxygen plasma. 
To fabricate the Schottky junctions, a layer of Ni (20 nm), capped with Au (20 nm) to prevent oxidation, was grown by electron beam evaporation at a room temperature at a base pressure of $\approx 10^{-6}$ Torr. 
Junctions with areas between $50\times100$ and $100\times300$ $\mu\text{m}^2$ were defined using UV-photolithography and ion beam etching. 
Further UV-photolithography, deposition and lift-off steps were done to create contact pads and finally wire bonding was carried out to connect the contacts to a chip carrier. 
Current-voltage measurements were performed using a Keithley 2410 SourceMeter in a cryostat under vacuum conditions ($10^{-6}$ Torr). 
All electrical measurements were done using three-terminal geometry with a voltage applied to the top Ni electrode. 
The measurements in the present work are performed on junctions of 20000 $\mu\text{m}^2$.
More details on the measurement techniques and device characterisation can be found in (\cite{goossens2018}). 
In light of practical applications and due to the attractive lowering of reverse bias current with increasing temperatures (\cite{susaki2007,goossens2018}), all measurements were done at room temperature. 

\subsection{Device Experimental Evaluation}

Biological synapses strengthen and weaken over time in a process known as \textit{synaptic plasticity}, which is thought to be one of the most important underlying mechanisms enabling learning and memory in the brain.
For a device to be useful as a neural network component, its resistance should thus be controllable through an external parameter - such as voltage pulses - so that it may take on a range of values (within a certain window).
The as-fabricated Ni/Nb-doped SrTiO$_3$ memristive devices showed stable and reproducible resistive switching without any electroforming process.

Hence, to investigate the effect of applying pulses across the interface, we conducted a series of measurements in which a device was subjected to a SET voltage of $+1$ V for 120 s to bring it to a low resistance state and reduce the influence of previous measurements. 
Then, 25 RESET pulses of negative polarity ($-4$ V) were applied and the current was read after each pulse. 
A negative read voltage was chosen because significantly larger hysteresis is observed in reverse bias compared to forward bias. 
Because of differences in the charge transport mechanisms under forward and reverse bias, a much smaller current flows in this regime: to ensure the measured current is sufficiently large to be read without significant noise levels, a reading voltage of $-1$ V was chosen. 
The RESET pulse amplitude was varied from $-2$ to $-4$ V. 
This procedure was repeated several times for each amplitude sequentially. 
The pulse widths were $\approx$1 s. 

Next, we performed a set of measurements in which devices were SET to a low resistance state by applying $+1$ V for 120 s before applying 10 RESET pulses of -4 V to bring devices to a depressed state. 
This was followed by applying a series of potentiation pulses ranging from $+0.1$ V to $+1$ V.

\subsection{Nengo Brain Maker Framework}

The aforementioned devices were integrated into a simulated spiking neural network (SNN); this was implemented using the Nengo Brain Maker Python package (\cite{bekolay2014nengo}), which represents information according to the principles of the Neural Engineering Framework (NEF) (\cite{eliasmith2003neural}).
Nengo was chosen because its two-tier architecture enables it to run on a wide variety of hardware backends - including GPUs, SpiNNaker, Loihi - with minimal changes to user-facing code, opening up the possibility of running models on a variety of neuromorphic hardware (\cite{10.1109/jproc.2014.2304638,davies2018loihi}).
The results of our current work are framework-independent and would apply to any computational model congruent to the one we simulated in Nengo.

In Nengo, information is represented by real-valued vectors which are \textit{encoded} into neural activity and \textit{decoded} to reconstruct the original signal (\cite{10.3389/fninf.2013.00048}).
The representation is realised by \textit{neuronal ensembles}, which are collections of neurons representing a vector; the higher the number of neurons in the ensemble, the better the vector can be encoded and decoded/reconstructed.
Each neuron $i$ in an ensemble (which in our case are leaky integrate-and-fire (LIF) neurons) has its own \textit{tuning curve} that describes how strongly it spikes in response to particular inputs (\textit{encoding}).
The collective spiking activity of the neurons in an ensemble represents the input vector, which can be reconstructed in output by applying a temporal filter to each spike train and linearly combining these individual contributions (\textit{decoding}).
Thus, each neuronal ensemble represents a vector through its unique neuronal activation pattern through time.

Ensembles are linked by \textit{connections} which transform and transmit the encoded vector represented in the \textit{pre-synaptic ensemble} by means of a \textit{weight matrix}.
The \textit{post-synaptic ensemble} will thus be trying to render a \textit{transformation} of the vector represented in the pre-synaptic neuronal ensemble; in the default case Nengo attempts offline optimisation to find a series of optimal weights to approximate the function along the connection.

\textit{Learning rules} can be applied to the connection between two ensembles in order to iteratively modify its weighting and thus find the correct transformation in an \textit{online} manner.

\subsection{Simulated Model}
\label{subsect:simulatedmodel}

\begin{figure}
    \centering
     \includegraphics[width=\textwidth]{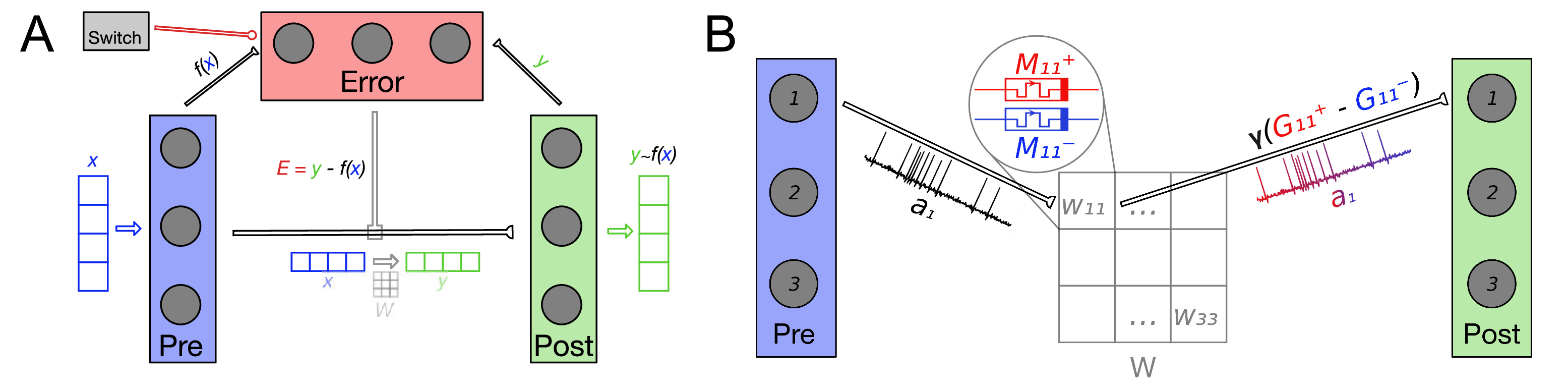}
    \caption{
    \textbf{(A)} The Error neuronal ensemble calculates the difference $\vec{E}=\vec{y}-f(\vec{x})$ between the post-synaptic $\vec{y}$ and transformed pre-synaptic $f(\vec{x})$ representations in order to inform the mPES learning rule.
    These connections are not implemented using simulated memristors and are calculated by Nengo.
    mPES tunes the weights $W$ on the connection between the pre- and post-synaptic ensembles with the overall goal of reducing the error $E$.
    $W$ transforms $\vec{x}$ into $\vec{y} \approx f(\vec{x})$ across the synaptic connection and, with training, the post-synaptic ensemble's representation improves, becoming closer to the transformed pre-synaptic one: $\vec{y} \rightarrow f(\vec{x})$.
    The Switch ensemble is activated to inhibit the Error ensemble and stop learning.
    NB: the input $\vec{x}$, error signal $\vec{E}$, and output $\vec{y}$ can have different dimensionality, which - in turn - is independent from the number of neurons used to represent them.
    In our current work, all three signals have the same dimensionality.
    \textbf{(B)} In this example, the pre- and post-synaptic neuronal ensembles both have size 3 and are fully-connected so the weight matrix $W$ has $3 \times 3$ entries.
    Each pair of pre- and post-synaptic neurons $(i,j)$ is linked by a simulated synapse.
    Each synapse's weight $W_{ij}$ is given by the scaled difference in normalised conductances $\gamma(G_{ij}^+ - G_{ij}^-)$ of its differential synaptic pair of memristors $M_{ij}^+$ and $M_{ij}^-$.
    As an example, the figure focuses on how the neurons $(1,1)$ in the pre- and post-synaptic ensembles are connected by a synaptic weight $W_{11}$; the neural activity $a^{\text{pre}}_1$ of the pre-synaptic neuron is combined with the synaptic weight $W_{11}$ to give the input to the first post-synaptic neuron.
    The activations of the post-synaptic neurons, in turn, define the signal represented in post-synaptic ensemble.
    }
     \label{fig:simulatedmodel}
 \end{figure}

The Nengo model used to evaluate the usage of our memristive devices in a neuromorphic learning setting, consisted of three neuronal ensembles: pre- and post-synaptic, and one for calculating an error signal $\vec{E}$. 
The pre- and post-synaptic ensembles were linked by a connection which resulted in a fully-connected topology between their constituent neurons; this is equivalent to the physical arrangement of memristors into \textit{crossbar arrays} (\cite{10.1038/s41563-019-0291-x}) used in many previous works to realise efficient brain-inspired computation (\cite{10.1038/s42256-018-0001-4,10.1109/tnnls.2013.2296777}).
The specific topology, which is crucial in enabling the model to learn, is shown in Fig. \ref{fig:simulatedmodel}A.

The pre-synaptic ensemble had as input signal a $d$-dimensional vector $\vec{x}$ and the connection between the ensembles was tuned in order for the post-synaptic ensemble to represent a transformation/function $f: \vec{x} \rightarrow \vec{y}$ of the vector encoded into the pre-synaptic neuronal population.
This time-varying input signal consisted of either $d$ uniformly phase-shifted sine waves described by
\begin{equation}
    x = \sin(\frac{1}{4} 2 \pi t + i \frac{2\pi}{d}), i \in [0,d]
\end{equation}
or of $d$ white noise signals low-filtered by a 5 Hz cutoff.
The latter signal was generated by using the \texttt{nengo.processes.WhiteSignal} class and is naturally periodic but, in order to be able to test network on unseen data, the period was chosen to be double the total simulation time of 30 s.


The third neuronal ensemble, dedicated to calculating the error signal, was connected to the pre- and post-synaptic ensembles using standard Nengo connections whose weights were pre-calculated by the framework.
The connection from the post-synaptic ensemble implemented the identity function - simply transferring the vector - while the one from the pre-synaptic ensemble calculated $-f(\vec{x})$ with $f$ being the transformation of the input signal we sought to teach the model to approximate.
The connection from pre to error ensembles did indeed already represent the transformation $f(\vec{x})$ that the model needed to learn, but the optimal network weights for this connection were calculated by Nengo before the start of the simulation based on the response characteristics of the neurons in the connected ensembles.
Therefore, the weights realising the transformation on this connection were not learned from data and did not change during the simulation.
Given these connections from pre and post ensembles, the error ensemble represented a $d$-dimensional vector $\vec{E}=\vec{y}-f(\vec{x})$ which was used to drive the learning in order to bring the post-synaptic representation $y$ as close as possible to the transformed pre-synaptic signal $f(\vec{x})$, the \textit{ground truth} in this \textit{supervised learning} context.  

The learning phase was stopped after 22 s of simulation time had passed; this was done by activating a fourth neuronal ensemble which had strong inhibitory connections to the error ensemble.
Suppressing the firing of neurons in the error ensemble stopped the learning rule on the main synaptic connection from modifying the weights further and thus let us test the quality of the learned representation in the remaining 8 s of simulation time.
An example simulation run is shown in Supplementary Fig. S2, S3.


In separate experiments, the transformation $f$ for the model to learn was either $f(\vec{x})=\vec{x}$ or $f(\vec{x})=\vec{x}^2$.
Learning the square function was expected to be a harder problem to solve as $f(\vec{x})=\vec{x}^2$ is not an \textit{injective} function; that is, different inputs can be mapped to the same output ($(-1)^2=1^2=1$, for example).
Even learning the identity function $f(\vec{x})=\vec{x}$ is not a trivial problem in this setting, as the pre- and post-synaptic ensembles are not congruent in how they represent information, even when they have the same number of neurons.
Each of the ensembles' neurons may respond differently to identical inputs, so the same vector decoded from the two ensembles will be specified by different neuronal activation patterns.
Therefore, the learning rule has to be able to find an optimal mapping between two potentially very different vector spaces even when the implicit transformation between these is the identity function.

The quality of the learned representation was evaluated on the last 8 seconds of neuronal activity by calculating the mean squared error (MSE) and Spearman correlation coefficient $\rho$ between the ground truth given by the transformed pre-synaptic vector $f(\vec{x})$, and the $\vec{y}$ vector represented in the post-synaptic ensemble.

At a lower level, each artificial synapse - one to connect each pair of neurons in the pre- and post-synaptic ensemble - in our model was composed of a ``positive'' $M^+$ and a ``negative'' $M^-$ simulated memristor (see Fig. \ref{fig:simulatedmodel}B).
The mapping of the current resistance state $R^\pm$ of a pair of memristors to the corresponding synaptic network weight $\omega$ was defined as:

\begin{align}
    \omega &= \gamma \left[ \left( \frac{\frac{1}{R^+} - \frac{1}{R_1}}{\frac{1}{R_0} - \frac{1}{R_1}  } \right) - \left( \frac{\frac{1}{R^-} - \frac{1}{R_1}}{\frac{1}{R_0} - \frac{1}{R_1}  } \right) \right] \nonumber \\
    \label{eq:resistance2conductance}
    &= \gamma \left[ \left( \frac{G^+ - G_0}{G_1 - G_0  } \right) - \left( \frac{G^- - G_0}{G_1 - G_0  } \right) \right] 
\end{align}
Note that the relationship between conductance and resistance was specified by $G=\frac{1}{R}$; thus, when normalising, the maximum conductance $G_1$ was given by the inverse of the minimum resistance $\frac{1}{R_0}$, and vice-versa.
So, the network weight of a synapse at each timestep was the result of the difference between its current memristor conductances $G^\pm$, which were normalised in the range $[0,1]$, and then multiplied by a gain factor $\gamma$. 

To simulate device-to-device variation and hardware noise we elected to introduce randomness into the model in two ways.
Firstly, we initialised the memristor resistances were initialised to a random high value in the range $[10^8 \, \Omega \pm 15\%]\, $ in order to ensure than all the weights were not $0$ at the start of the simulation.
This value was chosen on the basis of the experimental measurements shown in Fig. \ref{fig:experimentaldetails}, it being a high-resistance state the devices converged to after the application of -4 V RESET pulses.
We can therefore imagine the simulated memristors being brought to their initial resistances by the application of several -4 V RESET pulses prior to the start of the training phase.
As each entry in weight matrix $W$ was given by the difference in normalised conductances of a pair of memristors, initialising them all to the same resistance $10^8 \, \Omega$ would have lead them all to necessarily be equal to $0$, as per Eq. \ref{eq:resistance2conductance}.
In Machine Learning parlance, the act of assigning a random initial value network parameters is known as ``symmetry breaking''.
Secondly, we perturbed, via the addition of $15\%$ Gaussian noise, the parameters in Eq. \ref{eq:memristorlaw} used to calculate the updated resistance states of the simulated devices.
This was done independently at every timestep, and for every simulated memristive device.

\subsection{Learning Rule}
\label{subsec:learningrule}

\begin{figure}
    \centering
    \includegraphics[width=0.5\textwidth]{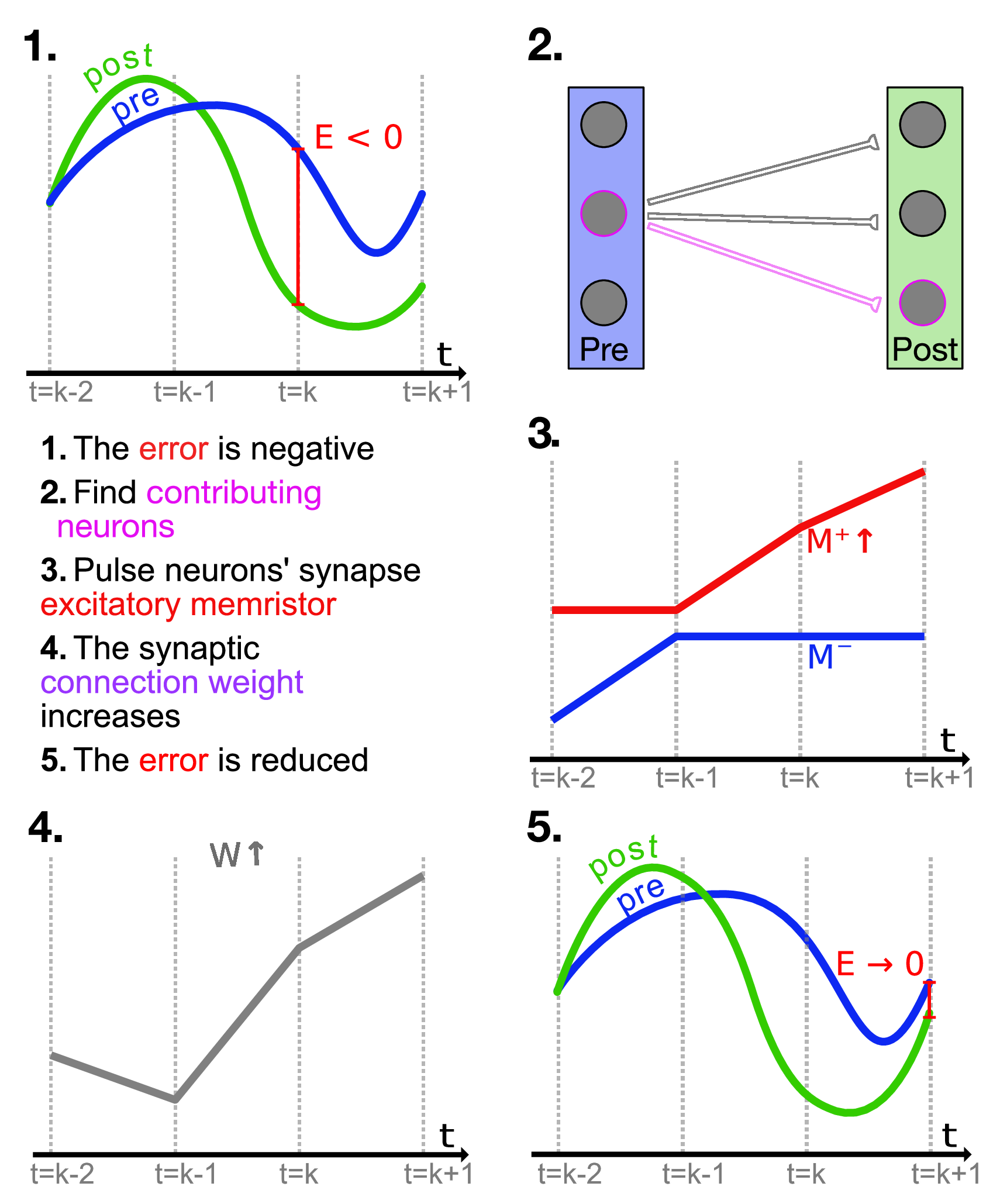}
    \caption{High level example (not to scale) of how the mPES learning rule minimises the error $\vec{E}$ between the pre- and post-synaptic representations.
    At time $t=k$ the neurons contributing to making the relevant component in error $\vec{E}$ negative are found, and the weights connecting them are adjusted accordingly; this example focuses on one such synapse.
    The error is negative so the synapse of the identified neurons is facilitated by pulsing its positive memristor $M^+$ to increase its conductance.
    The facilitation increases the future response of the post-synaptic neuron and this is conducive to reducing the error in the following timesteps.
    NB: The process is analogous for the PES learning rule, with the exception of step \textbf{3.}, as PES operates directly on the synaptic weights $W$.
    }
    \label{fig:learningrule}
\end{figure}

\begin{figure}
    \centering
    \includegraphics[width=0.5\textwidth]{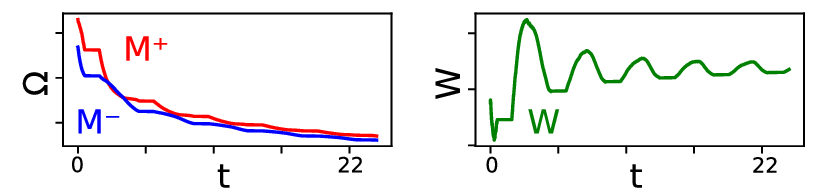}
    \caption{The left panel shows the evolution of the memristors' resistances in one differential synaptic pair during the learning part of the simulation.
    The positive memristor $M^+$ in the pair in plotted in red, the negative memristor $M^-$ in blue.
    The right panel shows the corresponding network weight $W$ whose value is given by applying Eq. \ref{eq:resistance2conductance} to the resistances in the differential synaptic pair.
    $W$ is unit-less and varies in the range $[0,\gamma]$.}
    \label{fig:resistanceweight}
\end{figure}

The process through which learning is effected in an artificial neural network is the iterative, guided modification of the network weights connecting the neurons.
To this end, we designed a neuromorphic learning rule that enacted this optimisation process by applying SET pulses to our memristive devices.

In order to learn the desired transformation from the pre-synaptic signal, we iteratively tuned the model parameter matrix $W$, which represented the transformation across the synaptic connection between the pre- and post-synaptic ensembles.
This was done by adjusting the resistances $R^\pm$ of the memristors in each differential synaptic pair through the simulated application of $+0.1$ V SET learning pulses.
At each timestep of the simulation, the resistances were adjusted using a modified version of the supervised, pseudo-Hebbian prescribed error sensitivity (PES) learning rule (\cite{10.1371/journal.pone.0022885}); this extension of PES to memristive synapses will be referred to as \textit{mPES} from here on.

The PES learning rule accomplishes online error minimisation by applying at every timestep the following equation to each synaptic weight $\omega_{i j}$:
\begin{equation}
\label{eq:pes}
    \Delta \omega_{i j}=\kappa \alpha_{j} e_{j} \vec{E} a_{i}
\end{equation}
where $\omega_{ij}$ is the weight of the connection between pre-synaptic neuron $i$ and post-synaptic neuron $j$, $\kappa$ is the learning rate, $\alpha_{j}$ the gain factor for neuron $j$, $e_{j}$ the encoding vector for neuron $j$, $\vec{E}$ the global $d$-dimensional error to minimise, and $a_{i}$ the activity of pre-synaptic neuron $i$.
$\alpha_j$ and $e_j$ are calculated for each neuron in the post-synaptic ensemble by the Nengo simulator such that:
\begin{equation}
    a_{j}=G\left(\alpha_{j} e_{j} \cdot \vec{x}+b_{j}\right)
\end{equation}
with $G$ the neuronal non-linearity, $\vec{x}$ the input vector to the ensemble, and $b_{j}$ a bias term.

The factor $\vec{\epsilon} = \alpha_{j} e_{j} \vec{E}$ is analogous to the \textit{local error} in the backpropagation algorithm widely used to train artificial neural networks (\cite{lecunbackprop}).
The overview shown in Fig. \ref{fig:learningrule} applies to both the PES and mPES learning rules, with the exception of step 3 that is characteristic to mPES; in this high-level depiction we can see how the neuronal pairs $i,j$ whose $\omega_{ij}>0$ are the ones which contribute to the post-synaptic representation being lower than the pre-synaptic signal and should thus be the ones potentiated in order to bring the post-synaptic representation closer to the pre-synaptic one.
The converse is also true, so the synapses whose corresponding $\omega_{ij}<0$ should be depressed in order to reduce $\vec{E}$.
Thus, the neurons in the post-synaptic ensemble that are erroneously activating will be less likely to be driven to fire in the future, changing the post-synaptic representation towards a more favourable one.

mPES is a novel learning rule developed to operate under the constraints imposed by the memristive devices and - essentially - behaves as a discretised version of PES.
This discretisation is a natural consequence of the models it operates on not having ideal continuous network weights but instead, by these being given by the state of memristors, which were updated in a discrete manner through voltage pulses.
The gradual change in the weights $W$ between pre- and post-synaptic neuronal ensembles in the model can be classified as a \textit{self-supervised learning} optimisation process, as the learning was informed by the error $\vec{E}$ vector calculated by the third neuronal ensemble - which was also a part of the model.
Our learning rule had to be able to operate in a very restricted and stochastic environment given that we only supplied $+0.1$ V SET potentiation pulses to our memristive devices during the online training process, and that we introduced uncertainty on the initial resistance and update magnitude of the memristors. 
Furthermore, at most one memristor in each synapse could have its resistance updated (either $M^+$ or $M^-$) and the exact result of this operation was also uncertain due to the fact that we injected noise into the parameters in Eq. \ref{eq:memristorlaw} governing the update.
Fig. \ref{fig:resistanceweight} shows how the resistances of the memristors in one such differential synaptic pair are modulated during training, and how the corresponding weight evolves over time.
An interesting characteristic stemming from the power-law memristance behaviour is that each SET pulse has a smaller effect than the one preceding it; the effect this has on the network weight is to help it converge to an optimum in a manner very similar to the cooling in \textit{simulated annealing} (\cite{kirkpatrick1983optimization}), as can be seen for $W$ in Fig. \ref{fig:resistanceweight}.
This kind of behaviour was seen for most - if not all - of the synapses in our simulations.

The learning rule is gated, pseudo-Hebbian, and uses discrete updates and local information, therefore making it biologically plausible.
Our claim that the learning rule is local is substantiated by the fact that only information available to post-synaptic neuron $j$ is used when calculating the weight update $\Delta \omega_{ij}$.
The key point is that each neuron has a different random encoder and is therefore responsive to a different portion of the error vector space.
Each update is thus optimising a separate local error $\vec{\epsilon}$ which depends on the neural state of $j$; there is no collaboration between units towards optimising a global error function as in classical gradient methods. 
The algorithmic approach is akin to that presented in - for example - (\cite{10.3389/fnins.2018.00608}).
Specific to our current setup is also the fact that we have no \textit{hidden layers} in the network and therefore all information is necessarily local, there being no intermediaries.
The characterisation as a gated learning rule derives from the fact that when $\vec{E}=\vec{0}$ learning does not occur, while the pseudo-Hebbian element is given by the post-synaptic activity being represented indirectly by $\alpha_{j} e_{j}$ instead of explicitly by $a_j$.

mPES required the definition and tuning of a hyperparameter $\gamma$, which is analogous to the \textit{learning rate} found in most machine learning algorithms.
$\gamma$ represents the gain factor used to scale the memristors' conductances $G^\pm$ to network weights $W$ whose magnitude was compatible with the signals the model needs to represent.
A larger $\gamma$ made each pulse to the memristors have an increased effect on the network weights but, unlike a learning rate $\kappa$, the magnitude of each update on the underlying objects was not changed by $\gamma$, affecting only the translation to network weights.

Our learning rule also presented an \textit{error threshold} $\theta_\epsilon$ hyperparameter whose role was to guarantee that a very small error $E$ would not elicit updates on the memristors, given that this was seen to negatively impact learning.
At this stage, the value for the error threshold was experimentally determined but kept constant and considered a static part of the model rather than a hyperparameter.

The mPES learning rule went through the following stages at each timestep of the simulation, in order to tune the resistances of each memristor in every differential synaptic pair with the overall goal of minimising the global error $E$:
\begin{enumerate}
    \item The \textit{global error} signal vector $\vec{E}$ was taken as input to the learning rule.
    \item A \textit{local error} $\vec{\epsilon}$ vector was calculated as $\vec{\epsilon} = -e_{j} \vec{E}$.
    $\vec{\epsilon}$ is a projection of the $\vec{E}$ onto each neuron's response and, consequently, encodes the contribution of each post-synaptic neuron to the global error $\vec{E}$ 
    (Note that the sign of $\vec{E}$ is inverted.)
    \item If no value in the $\vec{\epsilon}$ vector was greater than an experimentally determined \textit{error threshold} $\theta_{\vec{\epsilon}}=10^{-5}$ the updates in this timestep were skipped.
    The reason for this check was twofold.
    Firstly, if we didn't filter for local errors very close to 0, the learning performance would suffer.
    This is because a very small error may have no consequence in a continuous-update setting, as that where original PES is applied, but in our setup it could trigger a voltage pulse on the corresponding memristor and bring it to a new resistance state.
    As memristors move between discrete resistance states, such an update would end up having an outsized effect. 
    Secondly, inhibiting the error ensemble to terminate the learning phase did not completely shut off its activity so $\vec{E}$ would never be exactly $\vec{0}$ and the adjustments to the resistances would not stop, thus leading to the learning phase going on indefinitely.
    \item A \textit{delta} matrix $\Delta=-\vec{\epsilon} \otimes \vec{a^{\text{pre}}}$, whose entries were the equivalent to the PES $\Delta \omega_{i j}$ in Eq. \ref{eq:pes}, was calculated.
    Each entry in $\Delta$ has a one-to-one correspondence to a synapse in the model and encoded the participation of each pair of pre- and post-synaptic neurons $(i,j)$ to the error $\vec{E}$.
    Therefore, the sign of each entry $\Delta_{ij}$ determines if the corresponding synapse should be potentiated ($\Delta_{ij}>0$) or depressed ($\Delta_{ij}<0$).
    \item The $\vec{a^{\text{pre}}}$ vector of pre-synaptic activations' elements were discretised into binary values 0 and 1, as the only information of interest was if a pre-synaptic neuron had spiked, not the intensity of its activity.
    \item The update direction $V$ for each synapse was calculated as $V=\text{sgn}(\Delta)$, in order to determine which memristor in each pair needed adjusting:
    \begin{enumerate}
        \item The positive memristor $M^+_{ij}$ in a differential synaptic pair was pulsed when the corresponding term $V_{ij}>0$.
        \item The negative memristor $M^-_{ij}$ was pulsed when $V_{ij}<0$.
    \end{enumerate}
    This lead to a facilitation of synapses whose neurons $(i,j)$ had a positive participation to the error $\vec{E}$ and a depression of those for whom $\Delta_{ij}<0$.
    The synapses connecting neurons which ``pushed'' the post-synaptic signal $\vec{y}$ to be higher than the reference one $f(\vec{x})$ were depressed in order to reduce the future activation of these same post-synaptic neurons.
    The magnitude of change induced by each pulse was determined by the memristive devices' characteristic response, which in this work was given by Eq. \ref{eq:memristorlaw}.
    \item Finally, Eq. \ref{eq:resistance2conductance} was applied to each synapse in order to convert from $M^\pm$'s resistances to network weights $W$.
\end{enumerate}

\subsection{Optimisation Experiments}

In order to assess the effect that the gain $\gamma$ hyperparameter had on the learning capacity of the model, we set up a series of eight experiments, one for each combination of the three binary parameters: input function (sine wave or white noise), learned transformation ($f(\vec{x})=\vec{x}$ or $f(\vec{x})=\vec{x}^2$), and ensemble size (10 or 100 pre- and post-synaptic neurons).
In each experiment we ran 100 randomly initialised models for each value of $\gamma$ logarithmically distributed in the range $[10,10^6]$ and recorded the average mean squared error and Spearman correlation coefficient measured on the testing phase of each simulation, during the final eight seconds.

\subsection{Learning Experiments}

To assess the learning capacity of our model and learning rule (mPES) we compared them against a network using prescribed error sensitivity (PES) as learning algorithm and continuous synapses together with noiseless weight updates.
The default model with mPES used to generate our results was run with 15\% of injected Gaussian noise on both the memristor parameters and the initial resistances, as previously described.
To show that our model was not just learning the transformed input signal $f(\vec{x})$ but the actual function $f$, we tested both on the same signal that had been learned upon, and on the other.
In other words, the model learned the $f(\vec{x})$ transformations from both a sine wave and a white noise signal, using either 10 or 100 neurons, and the quality of the learning was then tested on both functions.
Both input learning and testing signals were 3-dimensional in all cases.

To have a quantitative basis to compare the two models, we measured the mean squared error (MSE) together with Spearman correlation $\rho$ between the pre- and post-synaptic representations in the testing phase of the simulation (corresponding to the last 8 of the total 30 s). 
We took PES as the ``golden standard'' to compare against, but also looked at what the error would have been when using the memristors without a learning rule modifying their resistances.
A lower mean squared error indicates a better correspondence between the pre- and post-synaptic signals, while a correlation coefficient close to $+1$ describes a strong monotonic relationship between them.
A higher MSE-to-$\rho$ ($\frac{\rho}{\text{MSE}}$) ratio suggests that a model is able to learn to represent the transformed pre-synaptic signal more faithfully.

\subsection{Other experiments}

We also tested the learning performance of the network for varying amounts of noise, as defining models capable of operating in stochastic, uncertain settings is one of the fundamental goals of the field on Neuromorphic Computing. 
To this end, we defined the simplest model possible within our learning framework (10 neurons, learning identity from sine input) and used it to evaluate the learning performance for varying amounts of Gaussian noise and for different values of the $c$ parameter in the resistance update Eq. \ref{eq:memristorlaw}.
We ran 10 randomly initialised models for each of 100 levels of noise - expressed as coefficient of variation $\frac{\sigma}{\mu}$ - in the range $[0\%,100\%]$ and reported the averaged MSE-to-$\rho$ ratio for each noise percentage.
The Gaussian noise was added in the same amount to the $R_0$, $R_1$, and $c$ parameters in Eq. \ref{eq:memristorlaw}. as well as to the initial resistances of the memristors.
The methodology we followed closely resembled that reported in (\cite{10.1109/tnano.2013.2250995}), but is should be noted that, in doing so, the magnitude of the uncertainty of each memristance update depends on the devices' current resistance state; it is presently unclear if this is the most physically correct method, but it suffices for the scope of this work.

Using the same methodology, we also ran a \textit{parameter search} for the exponent $c$ in Eq. \ref{eq:memristorlaw} in order to find the SET voltage magnitude best able to promote learning.
To realise this, we evaluated the MSE-to-$\rho$ ratio of randomly initialised models with $15\%$ noise for 100 values of $c$ uniformly distributed in the range $[-1,-0.0001]$.

\section{Results}
\label{sec:results}

\subsection{Device Experimental Evaluation}

\begin{figure}
    \centering
     \includegraphics[width=\textwidth]{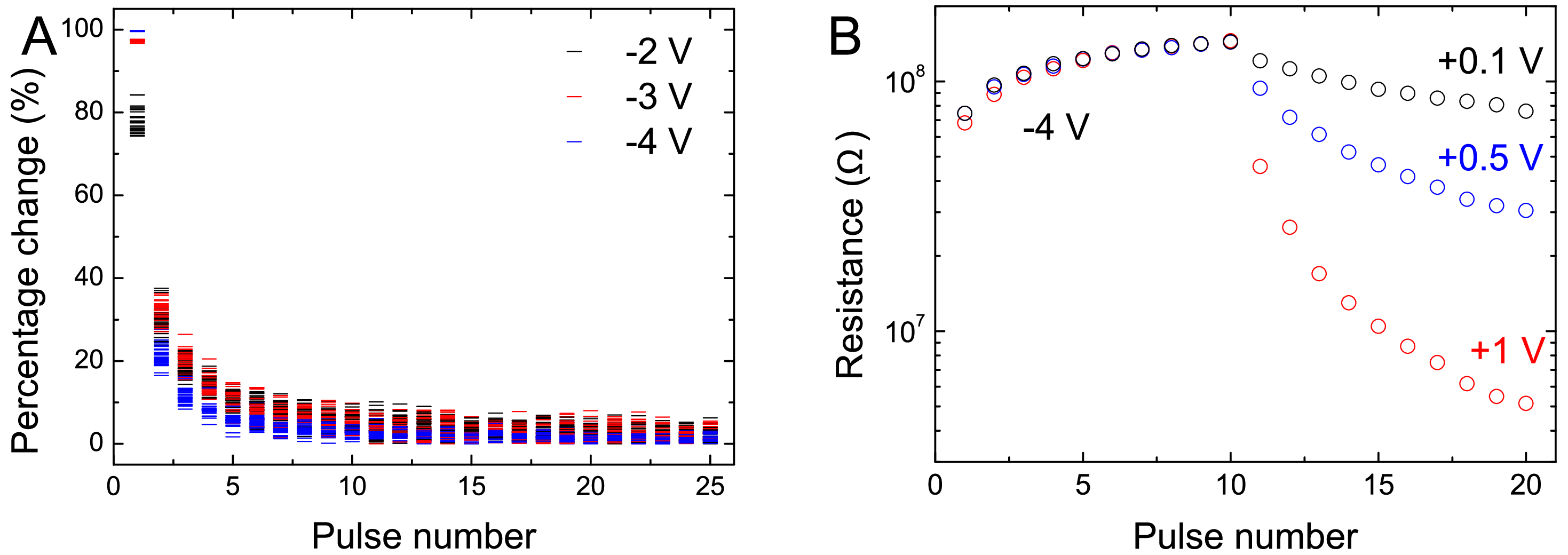}
     \caption{
     \textbf{(A)} Device response to the application of multiple RESET pulses of $-4$ V. Supplementary Fig. S1 shows the results for varying RESET amplitudes.
    \textbf{(B)} Device response to the application of multiple SET pulses of varying amplitudes of $+0.1$ V (black), $+0.5$ V (blue) and $+1$ V (red). 10 RESET pulses of -4 V are applied to increase the resistance before the application of the potentiation pulses.
    }
     \label{fig:experimentaldetails}
\end{figure}

We investigated the effect of applying pulses across the interface by administering a long SET pulse followed by 25 RESET pulses.
Due to the resistance's dependence on the memristor's previous history, there is some variation in the starting state, shown by the measurement at pulse number 0. Fig. \ref{fig:experimentaldetails}A shows the results for pulses of -4 V in amplitude. Results for other pulse amplitudes can be found in Supplementary Fig. S1.
The first pulse gave rise to the largest increase in resistance, with subsequent ones having a much smaller effect. 
The change in resistance quickly leveled off and the influence of subsequent pulses was significantly smaller. 
The application of a RESET pulse resulted in a switch to a high resistance state that depended strongly on the amplitude of the applied pulse, but not particularly on the number of pulses of that amplitude that were applied. 

We also explored the devices' response to SET voltages.
It should be pointed out that in Fig. \ref{fig:experimentaldetails}B the initial state (and hence also the large difference induced by the first RESET pulse) after applying a long SET voltage, which would correspond to pulse number 0, is not shown.
Note that different pulse amplitudes were chosen for SET and RESET pulses because the asymmetric nature of the charge transport results in much larger currents in forward bias (\cite{goossens2018}).
With the application of positive pulses, the device saw its resistance gradually decrease i.e., it was potentiated.
It is clear that in this case both the pulse amplitude and the number of applied pulses had a great impact on the resistance state of the device. 
The larger the amplitude of the SET pulse, the greater the induced difference to the resistance with each applied pulse. 
Interesting, compared to applying RESET pulses, the difference between the change induced by the first and later pulses was not severe. 
While 10 applied RESET pulses gave rise to close to saturated high resistance states, this was not observed when positive pulses were applied. 

\subsection{Device Memristive Response}

\begin{figure}
    \centering
     \includegraphics[width=\textwidth]{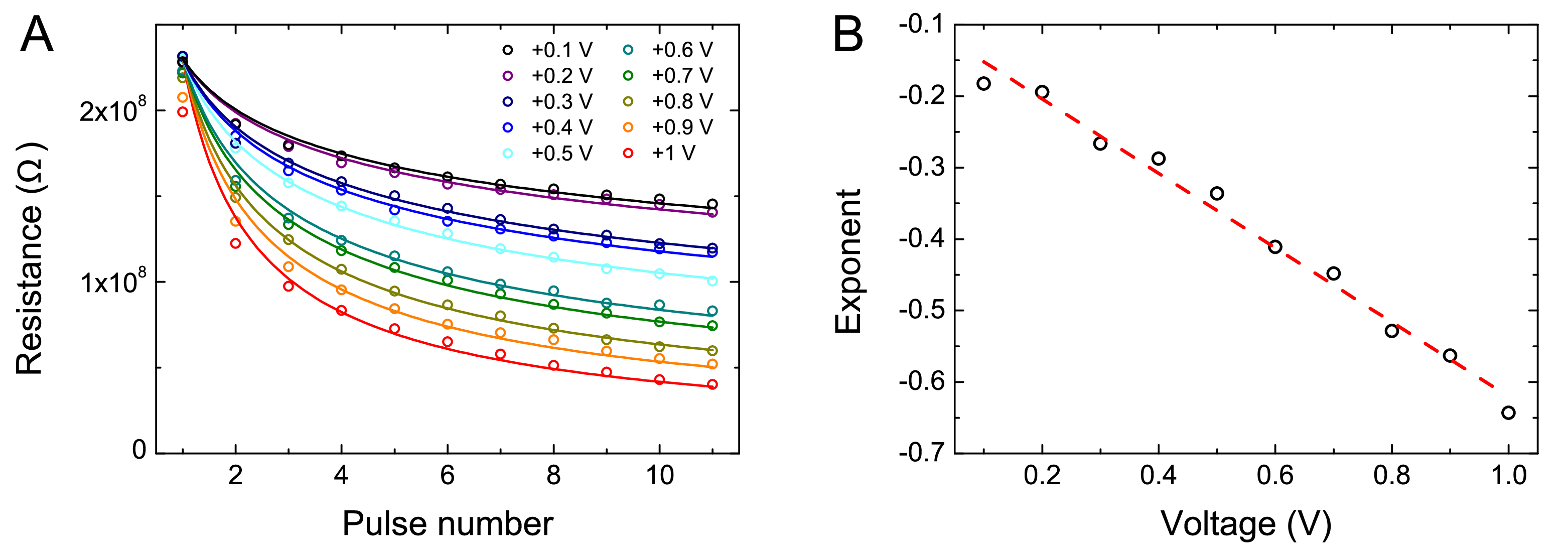}
     \caption{
     \textbf{(A)} Circles show the experimental data of device resistance after the application of multiple SET pulses with amplitudes varying from $+0.1$ V (top branch) to $+1$ V (bottom branch). Lines represent fits to Eq. \ref{eq:memristorgeneral}.
    \textbf{(B)} Black circles show the exponents extracted from the fits in (a) as a function of pulse voltage. The red line is a linear regression fit from which the $a$ and $b$ parameters are obtained.
    }
     \label{fig:deviceresponse}
 \end{figure}

In order to simulate our memristive devices in Nengo, we modelled their behaviour in response to SET pulses on the basis of the experimental measurements we carried out.

The device behaviour when applying a series of SET pulses, i.e. its \textit{forward bias} response, was seen to be well-described by an exponential equation of the form: 
\begin{equation} 
\label{eq:memristorgeneral}
R(n,V) = R_0 + R_1 n^{a+bV}
\end{equation} 
in which $V$ represented the amplitude of the SET pulse, $n$ the pulse number, $R_0$ the lowest value that the resistance $R(n,V)$ could reach, and $R_0+R_1$ the highest value.
One of the reasons we chose a fit of this form was because of the parallel we saw with the classic \textit{power law of practice}, a psychological and biological phenomenon by virtue of which improvements are quick at the beginning but become slower with practice.
In particular, skill acquisition has classically been thought to follow a power law (\cite{10.1080/00221325.1991.9914706,newell1981mechanisms}).

By solving Eq. \ref{eq:memristorgeneral} for $n$, we can calculate the pulse number from the current resistance $R(n,V)$ by: 
\begin{equation} 
n = \left( \frac{R(n,V) - R_0}{R_1} \right)^{\frac{1}{a+bV}} 
\end{equation} 
Parameters $a$ and $b$ were found based on the data measured by applying SET voltages between $+0.1$ V and $+1$ V to the memristive devices, as shown in Fig. \ref{fig:deviceresponse}A.
The exponents of the curves best describing the memristors' behaviour were then fitted with linear regression on the log-transformed pulse numbers and resistances, shown in Fig. \ref{fig:deviceresponse}B, in order to obtain a linear expression $a+bV$. 
This process yielded an estimated best fit for the memristor behaviour of: 
\begin{equation} 
\label{eq:memristorlawold} 
R(n,V) = 200 + 2.3 \times 10^8 n^{-0.093-0.53V} 
\end{equation} 
In our current work we always supplied SET pulses of $+0.1$ V so we could subsume the exponential term $a+bV$ into a single parameter $c$:
\begin{equation}
\label{eq:memristorlaw}
    R(n) = R_0 + R_1 n^{c} = 200 + 2.3 \times 10^8 n^{-0.093-0.53 \times 0.1} = 200 + 2.3 \times 10^8 n^{-0.146}
\end{equation}

\subsection{Optimisation Experiments}
\label{subsec:gammaexperiments}

\begin{table}[!t]
    \caption{
    Influence of the $\gamma$ hyperparameter on the learning performance, expressed as mean squared error (MSE), Spearman correlation coefficient ($\rho$), and MSE-to-$\rho$ ratio - measured in the testing phase - of the model for different combinations of ensemble size, input signal, and learned transformation.
    The bolded entries highlight the optimal value found for each statistic, and the best $\gamma$ chosen for each model.
    NB: the $\gamma$ chosen for the (\textbf{100 neurons, white,} $\mathbf{f(x)=x}$) model was $\gamma=10^4$ even though the $\frac{\rho}{\text{MSE}}$ was lower for $\gamma=10^3$.
    }
    \begin{minipage}{.5\linewidth}
        \centering
        \begin{tabular}{c|ccc}
        \multicolumn{4}{c}{{\textbf{10 neurons, sine,} $\mathbf{f(x)=x}$}} \\ 
            \toprule
            Gain $\gamma$ &  MSE & $\rho$ & $\frac{\rho}{\text{MSE}}$ \\ 
            \midrule
            $10^1$ & 0.3058          & 0.0422 & 0.1381          \\
            $10^2$ & 0.2737 & 0.3234 & 1.1814          \\
            $10^3$ & 0.1416          & 0.7997 & 5.6478          \\
            $\mathbf{10^4}$ & \textbf{0.1183}          & \textbf{0.8674} & \textbf{7.3303} \\
            $10^5$ & 0.2190 & 0.7670 & 3.5016          \\
            $10^6$ & 0.3177          & 0.6802 & 2.1408          \\
            \bottomrule
        \end{tabular}
    \end{minipage}
    \vspace{5pt}
    \begin{minipage}{.5\linewidth}
        \centering
        \begin{tabular}{c|ccc}
        \multicolumn{4}{c}{{\textbf{10 neurons, sine,} $\mathbf{f(x)=x^2}$}} \\ 
            \toprule
            Gain $\gamma$ &  MSE & $\rho$ & $\frac{\rho}{\text{MSE}}$ \\ 
            \midrule
            $10^1$ & 0.1710          & 0.0061          & 0.0356          \\
            $10^2$ & 0.1389          & 0.0364          & 0.2620          \\
            $10^3$ & \textbf{0.1188} & 0.1745          & 1.4696          \\
            $\mathbf{10^4}$ & 0.1346          & \textbf{0.2306} & \textbf{1.7129} \\
            $10^5$ & 0.2752          & 0.1161          & 0.4219          \\
            $10^6$ & 0.4189          & 0.0747          & 0.1784 \\
            \bottomrule
        \end{tabular}
    \end{minipage} 
    
    \begin{minipage}{.5\linewidth}
        \centering
        \begin{tabular}{c|ccc}
        \multicolumn{4}{c}{{\textbf{10 neurons, white,} $\mathbf{f(x)=x}$}} \\ 
            \toprule
            Gain $\gamma$ &  MSE & $\rho$ & $\frac{\rho}{\text{MSE}}$ \\ 
            \midrule
            $10^1$ & 0.1733          & 0.0107          & 0.0618          \\
            $10^2$ & 0.1631          & 0.1070          & 0.6563          \\
            $10^3$ & 0.1352          & 0.5313          & 3.9290          \\
            $\mathbf{10^4}$ & \textbf{0.1287} & \textbf{0.7599} & \textbf{5.9052} \\
            $10^5$ & 0.1830          & 0.7308          & 3.9923          \\
            $10^6$ & 0.2302          & 0.6136          & 2.6655  \\
            \bottomrule
        \end{tabular}
    \end{minipage} 
    \vspace{5pt}
    \begin{minipage}{.5\linewidth}
        \centering
        \begin{tabular}{c|ccc}
        \multicolumn{4}{c}{{\textbf{10 neurons, white,} $\mathbf{f(x)=x^2}$}} \\ 
            \toprule
            Gain $\gamma$ &  MSE & $\rho$ & $\frac{\rho}{\text{MSE}}$ \\ 
            \midrule
            $10^1$ & 0.0863          & 0.0030          & 0.0349          \\
            $10^2$ & \textbf{0.0752} & 0.0280          & 0.3722          \\
            $10^3$ & 0.0869          & 0.0882          & 1.0149          \\
            $\mathbf{10^4}$ & 0.1298          & \textbf{0.1542} & \textbf{1.1882} \\
            $10^5$ & 0.1831          & 0.1332          & 0.7276          \\
            $10^6$ & 0.2866          & 0.0936          & 0.3268   \\
            \bottomrule
        \end{tabular}
    \end{minipage} 
    
    \begin{minipage}{.5\linewidth}
        \centering
        \begin{tabular}{c|ccc}
        \multicolumn{4}{c}{{ \textbf{100 neurons, sine,} $\mathbf{f(x)=x}$ }} \\ 
            \toprule
            Gain $\gamma$ &  MSE & $\rho$ & $\frac{\rho}{\text{MSE}}$ \\ 
            \midrule
            $10^1$ & 0.3049          & 0.0357          & 0.1172          \\
            $10^2$ & 0.2740          & 0.3171          & 1.1571          \\
            $10^3$ & 0.1461          & 0.7886          & 5.3989          \\
            $\mathbf{10^4}$ & \textbf{0.1239} & \textbf{0.8741} & \textbf{7.0567} \\
            $10^5$ & 0.2341          & 0.7785          & 3.3254          \\
            $10^6$ & 0.3649          & 0.6649          & 1.8220            \\
            \bottomrule
        \end{tabular}
    \end{minipage} 
    \vspace{5pt}
    \begin{minipage}{.5\linewidth}
        \centering
        \begin{tabular}{c|ccc}
         \multicolumn{4}{c}{{ \textbf{100 neurons, sine,} $\mathbf{f(x)=x^2}$ }}  \\ 
            \toprule
            Gain $\gamma$ &  MSE & $\rho$ & $\frac{\rho}{\text{MSE}}$ \\ 
            \midrule
            $10^1$ & 0.1817          & 0.0065          & 0.0360          \\
            $10^2$ & 0.1439          & 0.0452          & 0.3144          \\
            $10^3$ & \textbf{0.1169} & 0.1511          & \textbf{1.2920} \\
            $\mathbf{10^4}$ & 0.1924          & \textbf{0.2143} & \textbf{1.1138} \\
            $10^5$ & 0.2710          & 0.1426          & 0.5261          \\
            $10^6$ & 0.4395          & 0.0843          & 0.1918   \\
            \bottomrule
        \end{tabular}
    \end{minipage} 
    
    \begin{minipage}{.5\linewidth}
        \centering
        \begin{tabular}{c|ccc}
        \multicolumn{4}{c}{{ \textbf{100 neurons, white,} $\mathbf{f(x)=x}$ }} \\ 
            \toprule
            Gain $\gamma$ &  MSE & $\rho$ & $\frac{\rho}{\text{MSE}}$ \\ 
            \midrule
            $10^1$ & 0.1761          & 0.0130          & 0.0736          \\
            $10^2$ & 0.1667          & 0.1001          & 0.6005          \\
            $10^3$ & 0.1413          & 0.5155          & 3.6488          \\
            $\mathbf{10^4}$ & \textbf{0.1219} & \textbf{0.7689} & \textbf{6.3070} \\
            $10^5$ & 0.1769          & 0.7292          & 4.1219          \\
            $10^6$ & 0.2225          & 0.6467          & 2.9062    \\
            \bottomrule
        \end{tabular}
    \end{minipage} 
    \vspace{5pt}
    \begin{minipage}{.5\linewidth}
        \centering
        \begin{tabular}{c|ccc}
        \multicolumn{4}{c}{{ \textbf{100 neurons, white,} $\mathbf{f(x)=x^2}$ }} \\ 
            \toprule
            Gain $\gamma$ &  MSE & $\rho$ & $\frac{\rho}{\text{MSE}}$ \\ 
            \midrule
            $10^1$ & 0.0790          & 0.0031          & 0.0389          \\
            $10^2$ & \textbf{0.0710} & 0.0255          & 0.3590          \\
            $10^3$ & 0.0928          & 0.0797          & 0.8583          \\
            $\mathbf{10^4}$ & 0.1255          & \textbf{0.1593} & \textbf{1.2694} \\
            $10^5$ & 0.1797          & 0.1373          & 0.7642          \\
            $10^6$ & 0.2844          & 0.0767          & 0.2696   \\
            \bottomrule
        \end{tabular}
    \end{minipage}
    \label{tab:1}
\end{table}

The gain factor $\gamma$ in Eq. \ref{eq:resistance2conductance} was analogous to the \textit{learning rate} $\kappa$ in Eq. \ref{eq:pes} - present in most machine learning algorithms - in that it defined how big an effect each memristor conductance update had on the corresponding network weight.

The rationale behind this experiment was to execute an equivalent of hyperparameter tuning - routinely done for artificial neural networks - as we had realised that $\gamma$ was homomorphous to a learning rate.
Therefore searching if there existed a ``best'' value of $\gamma$ was integral in helping mPES obtain good learning performance from the models.

The results for this experiment are shown in Table \ref{tab:1}, with the optimal value of the hyperparameter $\gamma$ that was found for each combination of factors highlighted in bold.
This ``best'' gain value was selected as the one giving the highest mean squared error-to-Spearman correlation coefficient ratio ($\frac{\rho}{\text{MSE}}$) across the various models.
In our case this value was $\gamma=10^4$, which would be the hyperparameter used in our subsequent experiments to evaluate the learning performance of our models in greater depth.
Most modern Machine Learning algorithms employ some form of scheduling of the learning rate in order to help convergence.
We had considered adding a decay to our gain factor but elected against this because, as previously stated, our eventual goal is to move our memristor-based learning model from simulation to physical circuit implementation and adding a schedule to our gain factor would entail a more complex CMOS logic.
It should also be noted that the memristors themselves naturally implement a decaying learning schedule as their memristance behaviour is described by a power law (Eq. \ref{eq:memristorgeneral}); each subsequent SET learning pulse will have a smaller effect on the resistance than the previous one.

\subsection{Learning Experiments}
\label{subsec:learningexperiments}

\begin{table}
\caption{
Mean squared error (MSE), Spearman correlation coefficient ($\rho$), and MSE-to-$\rho$ - averaged over 100 runs - for models trying to learn either $y=x$ or $y=x^2$ using PES or mPES as learning rule, from a 3-dimensional learning input signal.
The models are given by a combination of number of neurons in their ensembles, training input signal, and function learned.
Each model was tested on both kinds of testing input.
The results where mPES is run with no learning time allotted are also included in the ``No learning'' column.
}
\small
\centering
\begin{tabular}{@{}cccc|ccc|ccc|ccc@{}}
\toprule
\textbf{Neur.} & \textbf{Learn} $\mathbf{x}$ & $\mathbf{f(x)}$ & Test $x$ & \multicolumn{3}{c}{PES} & \multicolumn{3}{c}{mPES} & \multicolumn{3}{c}{No learning} \\
\multicolumn{4}{l}{} & MSE & $\rho$ & $\frac{\rho}{\text{MSE}}$ & MSE & $\rho$ & $\frac{\rho}{\text{MSE}}$ & MSE & $\rho$ & $\frac{\rho}{\text{MSE}}$ \\
\midrule
\multirow{8}{*}{10}  & \multirow{4}{*}{Sine}  & \multirow{2}{*}{$x$}   & Sine   & 0.2088 & 0.8283 & 3.9673 & 0.1283 & 0.8719 & 6.7957 & 0.5675 & 0.0511 & 0.0900     \\
                     &                        &                          & White  & 0.2800 & 0.5526 & 1.9735 & 0.1822 & 0.6268 & 3.4412 & 0.4044 & -0.0128 & -0.0317    \\
\cmidrule{3-13}
                     &                        & \multirow{2}{*}{$x^2$} & Sine   & 0.2048 & 0.2132 & 1.0411 & 0.1673 & 0.2032 & 1.2146 & 0.1849 & -0.0075 & -0.0406    \\
                     &                        &                          & White  & 0.1783 & 0.0390 & 0.2189 & 0.1322 & 0.0992 & 0.7499 & 0.1005 & 0.0153 & 0.1524     \\
\cmidrule{2-13}
                     & \multirow{4}{*}{White} & \multirow{2}{*}{$x$}   & Sine   & 0.1315 & 0.8708 & 6.6223 & 0.1712 & 0.8058 & 4.7077 & 0.8719 & 0.0222 & 0.0255     \\
                     &                        &                          & White  & 0.1515 & 0.8322 & 5.4928 & 0.1307 & 0.7719 & 5.9074 & 0.8021 & 0.0006 & 0.0008     \\
\cmidrule{3-13}
                     &                        & \multirow{2}{*}{$x^2$} & Sine   & 0.1561 & 0.2786 & 1.7854 & 0.1766 & 0.2024 & 1.1465 & 0.1676 & 0.0007 & 0.0042     \\
                     &                        &                          & White  & 0.1586 & 0.2266 & 1.4288 & 0.1247 & 0.1374 & 1.1013 & 0.0002 & -0.0009 & -5.6996    \\
\midrule
\multirow{8}{*}{100} & \multirow{4}{*}{Sine}  & \multirow{2}{*}{$x$}   & Sine   & 0.1385 & 0.8812 & 6.3601 & 0.1197 & 0.9421 & 7.8731 & 0.4208 & -0.0305 & -0.0724    \\
                     &                        &                          & White  & 0.1905 & 0.6810 & 3.5742 & 0.1847 & 0.7344 & 3.9758 & 0.2204 & -0.0179 & -0.0811    \\
\cmidrule{3-13}                     
                     &                        & \multirow{2}{*}{$x^2$} & Sine   & 0.0816 & 0.5956 & 7.3007 & 0.1316 & 0.4796 & 3.6441 & 0.2579 & -0.0322 & -0.1249    \\
                     &                        &                          & White  & 0.0997 & 0.1739 & 1.7445 & 0.1724 & 0.1215 & 0.7047 & 0.1222 & -0.0139 & -0.1140    \\
\cmidrule{2-13}
                     & \multirow{4}{*}{White} & \multirow{2}{*}{$x$}   & Sine   & 0.0176 & 0.9865 & 56.1002 & 0.0867 & 0.9614 & 11.0912 & 0.4653 & -0.0152 & -0.0327    \\
                     &                        &                          & White  & 0.0162 & 0.9784 & 60.2713 & 0.0654 & 0.9554 & 14.6141 & 0.2508 & 0.0186 & 0.0742     \\
\cmidrule{3-13}
                     &                        & \multirow{2}{*}{$x^2$} & Sine   & 0.0385 & 0.8384 & 21.7865 & 0.1405 & 0.5152 & 3.6670 & 0.2465 & -0.0008 & -0.0031    \\
                     &                        &                          & White  & 0.0248 & 0.7957 & 32.0590 & 0.1013 & 0.5604 & 5.5338 & 0.1122 & -0.0014 & -0.0129   \\
\bottomrule
\end{tabular}
\label{tab:2}
\end{table}

The results for the experiments comparing learning a function using PES or mPES are shown in Table \ref{tab:2}, where it can be immediately seen that our mPES learning rule is competitive with PES by noticing that the statistics encoding the quality of the learning, the mean squared error (MSE) and Spearman correlation coefficient ($\rho$), are consistently more favourable to mPES - or when not, very close to equal - across the spectrum of tested models.

\begin{figure}
    \centering
    \includegraphics[width=\textwidth]{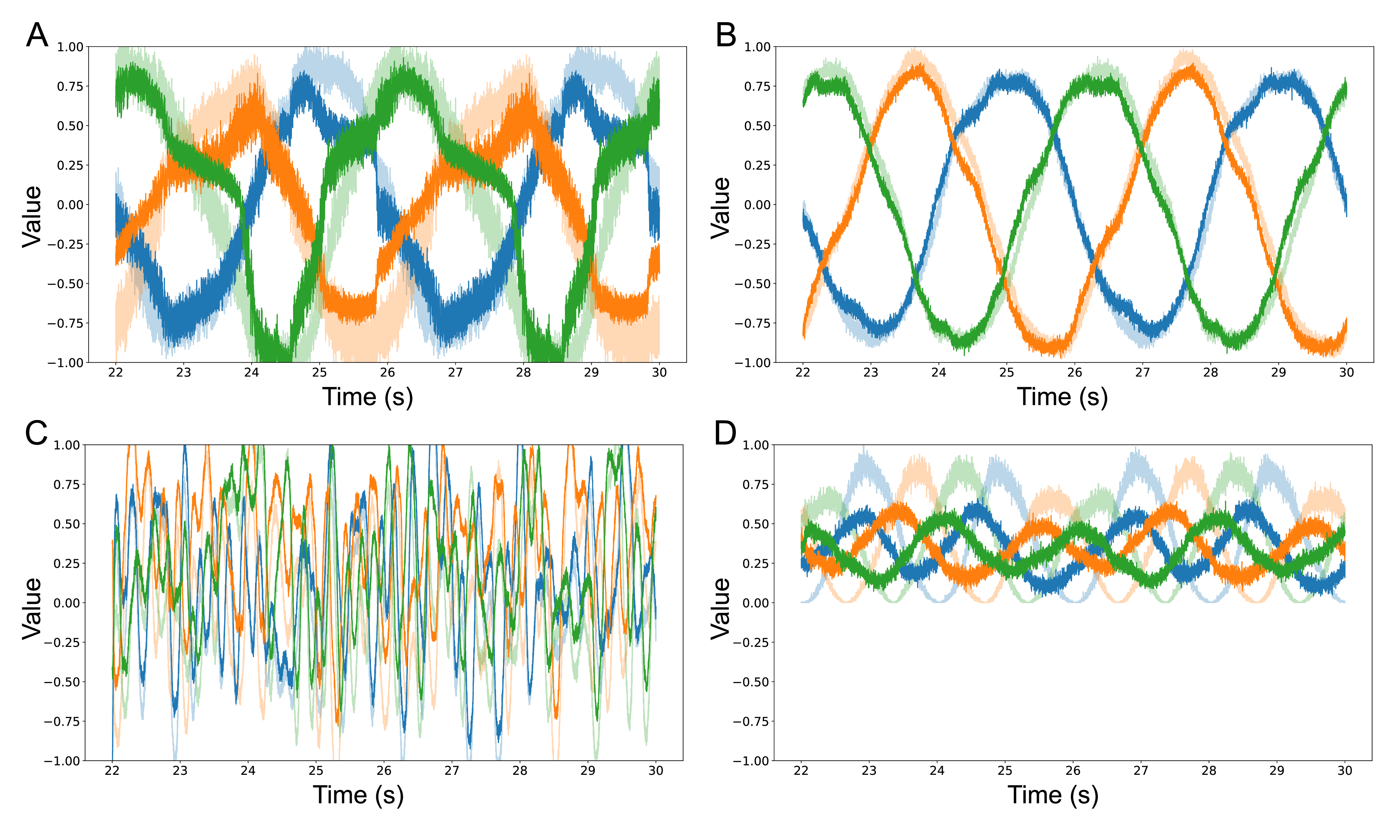}
    \caption{
    \textbf{(A)} Decoded output from pre- and post-synaptic ensembles of 10 neurons on a sine wave input, after being trained on a sine wave to learn to represent the identity function $f(\vec{x})=\vec{x}$.
    The quality of learning is $\frac{\rho}{\text{MSE}} = 16.3398$.
    \textbf{(B)} Decoded output from pre- and post-synaptic ensembles of 100 neurons on a sine wave input, after being trained on a sine wave to learn to represent the identity function $f(\vec{x})=\vec{x}$.
    The quality of learning is $\frac{\rho}{\text{MSE}} = 132.9364$.
    \textbf{(C)} Decoded output from pre- and post-synaptic ensembles of 100 neurons on a white noise input, after being trained on a white noise signal to learn to represent the identity function $f(\vec{x})=\vec{x}$.
    The quality of learning is $\frac{\rho}{\text{MSE}} = 10.5064$.
    \textbf{(D)} Decoded output from pre- and post-synaptic ensembles of 100 neurons on a sine wave input, after being trained on a sine wave to learn to represent the square function $f(\vec{x})=\vec{x}^2$.
    The quality of learning is $\frac{\rho}{\text{MSE}} = 12.2479$.
    }
    \label{fig:learningexperiments}
\end{figure}

Figures \ref{fig:learningexperiments}A-D show a selection of results from our simulations for various combinations of neuronal ensemble sizes, desired transformations $f$ and input signals.
The plots show the decoded output for the post-synaptic neuronal ensemble (filled colours), overlaid to the signal represented in the pre-synaptic ensemble (faded colours) during the testing phase of the simulation.
The x-axis shows the final 8 seconds of the simulation, while the y-axis is fixed to $[-1,+1]$ to show the value of the signals.
In all cases the training input and testing signals were 3-dimensional, with the first dimension plotted in blue, the second in orange, and the third in green.
These results were chosen to give an intuition on the outcome of learning from a qualitative point of view, given that mere statistics might not encode the notion of what an observer would deem a ``good'' fit.

Fig. \ref{fig:learningexperiments}A shows how pre, post and error ensembles of 10 neurons each, after the learning period, were able to learn the identity function from a 3-dimensional sine wave.
Increasing the model's ensembles to 100 neurons (see Fig. \ref{fig:learningexperiments}B) gave a much cleaner output that better approximated the pre-synaptic signal.
It should also be noted that the signal decoded from the smaller neuronal ensembles is much noisier, apart from having a worse average fit (i.e., has both higher MSE and lower $\rho$).
Fig. \ref{fig:learningexperiments}C shows that 100-neuron ensembles could learn the identity function after being trained on a white noise signal.
Finally, Fig. \ref{fig:learningexperiments}D indicates that neuronal ensembles of 100 neurons each could also learn to approximate the square from an input sine wave, with the caveat being that the quality of the learned transformation was noticeably lower - which is also reflected by the statistics in Table \ref{tab:2}.

An argument in support of reporting the notion of qualitative fit is given by analysing result in Table \ref{tab:2} for the models in Fig. \ref{fig:learningexperiments}B and Fig. \ref{fig:learningexperiments}D.
Here we see how these particular models perform similarly in regards to the measured mean squared error when trying to learn the identity and square function from a sine wave, but how the Spearman correlation coefficient is two times worse when learning the latter.
The answer to the dilemma can be found by looking at the qualitative fits on the testing part of the simulations (Fig. \ref{fig:learningexperiments}B and Fig. \ref{fig:learningexperiments}D) and noticing that, even though the pre- and post-synaptic signals are - on average - close in both cases, when trying to learn the square the post-synaptic signal is much noisier.
A noisy signal negatively impacts the correlation coefficient as the pre- and post-synaptic signals are nearly never moving in tandem, even when the oscillations are around the correct value.
This could be ascribed to the square being a harder function to learn than the identity, as already noted throughout, which could lead to a model being ``overpowered'', in the sense that its representational power is allotted to trying to learn the harder function, decreasing the resources available to represent the signal.

The sizes of the neuronal ensembles were not fine-tuned for compactness so it may well be possible to learn representations of similar quality using smaller neuronal ensembles.



\subsection{Other Experiments}

\begin{figure}
    \centering
    \includegraphics[width=\textwidth]{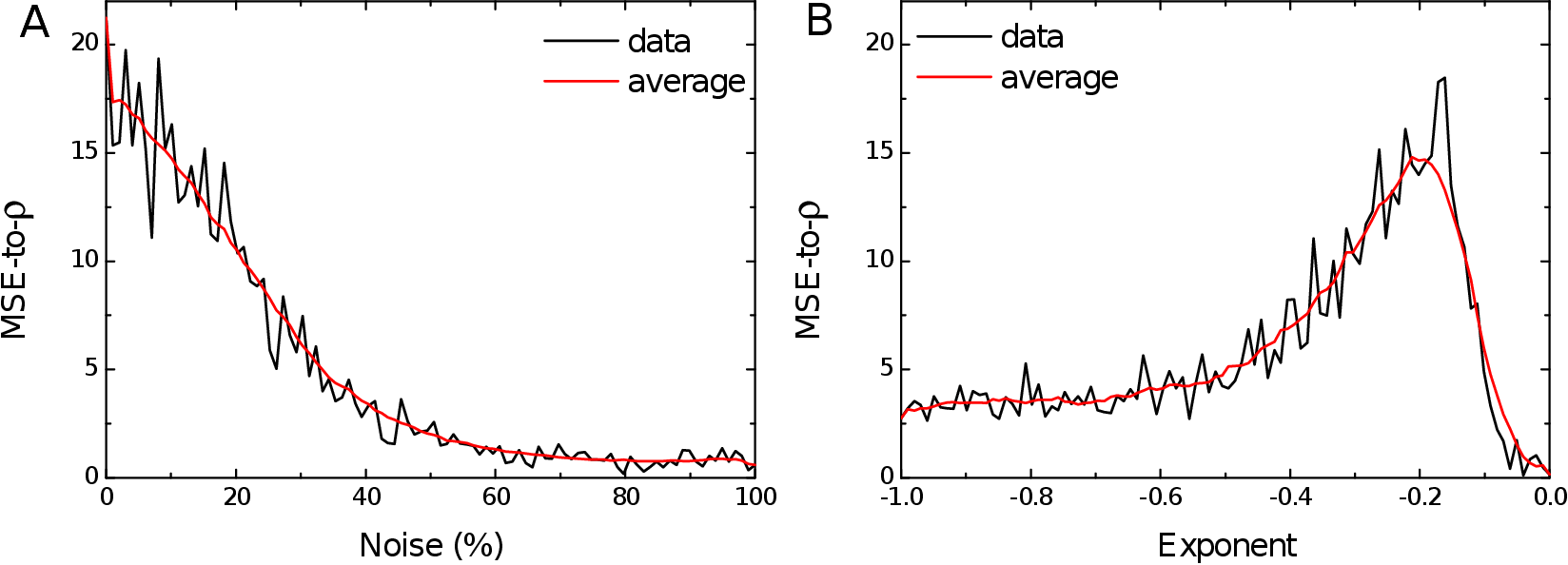}
    \caption{
    \textbf{(A)} The black line shows the MSE-to-$\rho$ ($\frac{\rho}{\text{MSE}}$) measured on the testing phase of the simulation as a function of the percentage of noise - expressed as coefficient of variation $\frac{\sigma}{\mu}$ of each parameter - injected into the model.
    The red line shows the rolling average for the measured MSE-to-$\rho$.
    \textbf{(B)} The black line shows the MSE-to-$\rho$ as a function of the magnitude of the exponent $c$ in Eq. \ref{eq:memristorlaw}. 
    The red line shows the rolling average for the measured MSE-to-$\rho$.
    }
    \label{fig:otherexperiments}
\end{figure}

\subsubsection{Increasing Noise Experiments}
\label{subsec:noiseexperiments}
As previously mentioned, the impact of noise is not linear across the resistance range, with memristors in a HRS seeing their response to pulses have a higher uncertainty.
This is due to the fact that we were introducing variation on the update equation itself and not on the resulting resistance state.

As can be seen in Fig. \ref{fig:otherexperiments}A, learning performance decreased for high levels of noise and this degradation was graceful, as the MSE-to-$\rho$ monotonically decreased towards $0$.
As previously stated: a lower MSE-to-$\rho$ ratio ($\frac{\rho}{\text{MSE}}$) indicates a model that cannot faithfully represent the transformed pre-synaptic signal $f(x)$.
Performance was well-maintained up until $\approx15\%$ of variation of the parameters and smoothly decreased until stabilising around $0$ for noise $>60\%$.
An example of how the memristor resistances behave in the presence of various amounts of noise is shown in Supplementary Fig. S4.

\subsubsection{Exponent Search Experiments}
The results for the exponent parameter search are shown in Fig. \ref{fig:otherexperiments}B and here we can see how strongly the performance of the model depends on the magnitude of the exponent in the power law (Eq. \ref{eq:memristorlaw}).
The quality of the fit is in direct correspondence with the value of $c$ up until $c\approx-0.162$, with a rapid decrease in learning performance thereafter.
The highest performance is found for $c \in [-0.17,-0.16]$.
Our choice of supplying learning SET pulses of $+0.1$ V implied $c=-0.146$ (Eq. \ref{eq:memristorlaw}), suggesting that a slightly higher tension could have led to better performance in the task at hand.

The decrease in MSE-to-$\rho$ for the smallest values of the exponent is most likely the result of the memristors having too small of a response for the training time supplied (which was 24 s in all out experiments) to be able to learn the function.
This is the equivalent to having too small a learning rate in a traditional machine learning model.



\section{Discussion}
Even with the restrictions imposed by the devices' physical characteristics, our experiments showed that we could enable the post-synaptic neuronal ensemble to well-represent transformations of the pre-synaptic signal by applying our novel mPES learning rule to synapses based on Nb:STO memristive devices.
This process leads to a set of memristor resistances that can be mapped to network weights, which transform the supplied input into the desired output by encoding the learned function.
Crucially, our results are not limited to our Nb:STO memristors but generalise to any memristive device whose memristance follows a power law.
We showed that our model could learn to approximate both the identity $f(\vec{x})=\vec{x}$ and square $f(\vec{x})=\vec{x}^2$ functions - the latter being a qualitatively harder problem to solve - with no theoretical reason limiting it to only representing these transformations.
Learning was effective both when the input signal was a periodic sine wave and also when it was random, as the case where the pre-synaptic ensemble was fed white noise, or was switched between learning and testing phases; this second result is important in that it shows that the model was able to learn the transformation independently of the regularity of the supplied training data and of the input signal itself.
It had already been suggested that the original PES learning rule, on which our mPES algorithm is based, would be able to learn arbitrary vector functions (\cite{10.1371/journal.pone.0022885, bekolay2010learning}).
The fact that our learning rule is derived from one already capable of approximating arbitrary functions leads further credence to our conclusion that our model is capable of learning any transformation of $n$-dimensional vectors.
Critically, as our model aimed to present strong neuromorphic credentials, these results are robust to the presence of noise on the updates and the initial state of the memristors' resistances.

The results in the learning experiments - during which we compared the quality of the representations in the pre- and post-synaptic ensembles - were competitive with the original PES learning rule even though the latter had the advantage of operating on ideal, continuous network weights, with no uncertainty on their updates.
Our mPES learning rule showed a post-training MSE-to-$\rho$ ratio that was consistently close to that of PES; the results were especially favourable when trying to learn the identity function $f(\vec{x})=\vec{x}$.
In general, the MSE-to-$\rho$ ratio was consistently lower for both algorithms when trying to learn the square transformation $f(\vec{x})=\vec{x}^2$ compared to the $f(\vec{x})=\vec{x}$ case, but the drop in performance for mPES was more marked.
These observations suggest that the square function is a harder problem to learn and that the gap in performance between PES and mPES when trying to learn this transformation could be bridged by using a higher number of neurons in conjunction with our learning rule.
Both learning rules exhibited higher performance when trained on a white noise input signal compared to a sine wave; we attribute this difference to the dissimilarity in how the input spaces are sampled but have no grounded theoretical explanations to explain this influence.
Most importantly, we did see a decline in learning performance when switching to a different input signal between learning and testing phases, but this effect was limited; this lends credence to the belief that our model is actually able to generalise and thus is learning the function $f$, not just the transformed input signal $f(\vec{x})$.

The evaluations on the learning performance for increasing amounts of noise were done with models composed of a very small number of neurons, which could potentially predispose the network to be more susceptible to the injected randomness; this suggests that using our memristive devices and learning rule can lead to robust performance. 
As previously reported, the model maintained its best performance up to a moderate amount of Gaussian noise on the initial memristor resistances and on the parameters of the update equation (Eq. \ref{eq:memristorlaw}).
The measured performance in the presence of increasing noise is less than that reported in previous work, but in our case we did not have a well-defined metric to measure performance against, for example recognition rate as in (\cite{10.1109/tnano.2013.2250995}).
Again we raise the question of what constitutes a ``good enough fit'', reiterating that - in our view - this is entirely dependent on the observer and the task at hand.
The results also suggest that noise aids the model to a certain degree, given that the act of initialising each network weight to a different value is, in Machine Learning terminology, known as ``symmetry breaking''; failing to do so can make a model difficult or impossible to train.
Therefore, the noise on the initial resistance state could be effectively ``breaking the symmetry'' of the weights.

It is notable that the learning performance was maintained even though our mPES learning rule had absolutely no knowledge of the magnitudes of the updates and that the only hyperparameter tuned was the gain $\gamma$, whose effect was analogous to a learning rate.
The learning rule was also only able to adjust the memristors towards lower resistance states, potentially making it harder to backtrack on any error.
We experimentally found, as reported in Table \ref{tab:1}, that a value of $\gamma = 10^4$ lead to the best learning performance - defined as balance between testing error and correlation MSE-to-$\rho$ ratio $\frac{\rho}{\text{MSE}}$ - for all models given by combinations of neuronal ensemble size, training input signal, and desired transformation.
The best values for mean squared error (MSE) and Spearman correlation coefficient ($\rho$) were both found for the same value of $\gamma = 10^4$ in all the instances where the models were instructed to learn the identity function $f(\vec{x})=\vec{x}$.
Instead, when the models were attempting to learn the square function $f(\vec{x})=\vec{x}^2$ there was a discrepancy between the values of $\gamma$ leading to the lowest MSE and to the highest $\rho$.
The two models learning the square from the regular sine wave input still had best $\rho$ for $\gamma = 10^4$, but showed the lowest error for a smaller gain factor of $\gamma = 10^3$; the combined MSE-to-$\rho$ statistic still indicated $\gamma = 10^4$ as the best choice for this class of model as the positive gains in correlation counterbalanced the higher error.
The models learning the square transformation from the random white noise input followed the same pattern as they also showed highest correlation for $\gamma = 10^4$ and required a smaller gain factor to exhibit the lowest error: $\gamma = 10^2$ in their case.
The measured $\frac{\rho}{\text{MSE}}$ for this last class of models was close for both $\gamma = 10^3$ and $\gamma = 10^4$ as the correlation $\rho$ did not suffer as much as when trying to learn the square from the sine wave input.
In the [100 neurons, sine, $f(\vec{x})=\vec{x}^2$] instance the best MSE-to-$\rho$ ratio was found for $\gamma = 10^3$ but, given that the result was very close to the one obtained with $\gamma = 10^4$, the latter was chosen for consistency with all other cases.

The magnitude of $\gamma$ was crucial to enable learning as it brought the normalised conductances calculated in Eq. \ref{eq:resistance2conductance} to a scale compatible with the model and the input signal.
The need to precisely tune the learning rate when using memristors presenting non-linear and asymmetric response has already been recognised (\cite{10.1109/ted.2015.2439635}) and this effect is somewhat confirmed here, as a too small or too large $\gamma$ completely disabled the learning capacity of the model, while a sub-optimal value hindered the learning performance.
Having to tune hyperparameters in order to achieve learning performance is, unfortunately, the norm in Machine Learning so it is not surprising that we ran into the same requirement.
The search was not fine-grained, but gave a good indication of the magnitude of $\gamma$ we could expect to lead to reasonable performance.
It could be argued that, in a neuromorphic setting, it is more advantageous to have a ``rough'' estimate for the hyperparameter - $\gamma = 10^4$ in our case - leading to ``robust'' learning in a variety of situations, rather than focusing on finding the optimal value for a specific setup.
Future work should focus on establishing a theoretical link between the model setup, including input signal characteristics and magnitude, and the expected range for finding a $\gamma$ leading to sufficiently robust learning performance.

The memristive response of our device had notable implications on its use as a substrate for learning (\cite{10.1109/ted.2015.2439635,10.1109/essderc.2016.7599680}).
There is contention regarding the best characteristics that devices' memristance should have in order to effectively support learning.
Recent works highlight how there is currently no consensus on if a linear response (\cite{10.1063/1.5108899}) or a non-linear (\cite{10.1088/1361-6528/aae81c,10.1038/s41598-018-25376-x}) behaviour of memristive synapses is more advantageous.
On our side, we can add to the discourse by pointing out that our learning experiments (results in Table \ref{tab:2}) showed that our mPES algorithm, modulating the resistance of non-linear Nb:STO memristors, was competitive with the original PES learning rule, which operated on linear, ideal synaptic weights.
Fig. \ref{fig:resistanceweight} adds some intuition to why a power-law update would be beneficial for learning compared to a linear synaptic response: the decreasing effect of each training pulse effectively mimics the cooling in simulated annealing and therefore helps the weight to converge to an optimal value.

In our current work we only used SET pulses of $+0.1$ V while negative RESET pulses were largely ignored as the effect of sequential negative pulses was seen to quickly diminish with pulse number. 
Our initial choice to apply positive learning pulses of $+0.1$ V implied having an exponent $c=-0.146$ for the power law governing the memristor updates, which put the model close to the region exhibiting best learning performance.
From a physical point of view, the magnitude of the SET pulses could be set very low as the current-voltage characteristics of our memristors are continuous, without any step-like features reminiscent of a switching threshold voltage.
Hence, in principle we expect even small pulse magnitudes to influence the resistive state of the device. 
In addition, by utilising a read voltage of a different polarity than that of the SET pulses, we created an asymmetry between the writing and reading processes that made it possible to use very small voltage amplitudes for the SET pulses.
Thus it should be possible, in principle, to use physical voltage pulses as small as needed to guarantee the best learning performance.
Different mechanisms govern charge transport in forward and reverse bias; this leads to asymmetries in current-voltage characteristics and may also result in differences between the SET and RESET processes. 

We focused on the forward bias regime for learning, by only supplying positive SET pulses to either the $M^+$ or $M^-$ memristor in each differential synaptic pair.
The disadvantage of such an approach is that long training regimes could potentially lead to the memristors' resistances to saturate.
That is, using only positive pulses progressively brings the devices to a low resistance state with no way to backtrack.
This issue could be alleviated by a careful pulsing scheme where memristor close to saturating were ``re-initialised'' to a high resistance state by supplying RESET voltage pulses, at the cost of losing some precision in the represented network weights (\cite{10.1109/ted.2015.2439635}).
This could be done by re-initialising both memristors in a pair to their initial resistances with RESET pulses and then supplying a number of SET pulses proportional to the magnitude of the weight before the reset, on either the positive of negative memristor depending on the polarity of the weight.
After the resetting, a highly positive weight would thus see its memristor $M^+$ receiving more pulses than a lightly positive one, as a moderately negative weight might have its memristor $M^-$ receive fewer SET pulses than a highly negative one.
This approach could have the disadvantage of requiring quite a complex meta-architecture, but it is possible that the current learning approach may already be robust enough to deal gracefully with the resetting of a small percentage of its synapses.


Another path that could improve the learning performance could be to grade the number, the duration, or the magnitude of pulses applied to each memristor, based on the participation of its pre- and post-synaptic neurons to the global error.
This could prove to be a less demanding undertaking than shaping the pulses based on the memristors' current state, while still resulting in positive gains to learning performance.


\section{Conclusions}
In this work we fabricated memristive devices based on Ni/Nb-doped SrTiO$_3$ and found that their memristance followed a power law. 
These memristive devices were used as the synaptic weight element in a spiking neural network to simulate, to our knowledge, one of the first models of this kind capable of learning to be a universal function approximator.
The performance was tested in the Nengo Brain Builder framework by defining a simple network topology capable of learning transformations from multi-dimensional, time-varying signals. 
The network demonstrated good learning performance with robustness to moderate noise levels of up to $\approx$15$\%$, showing that this class of memristive devices is apt to being used as component of a neuromorphic architecture.
It is worth restating that the network weights were found using only discrete updates to the memristors, based on knowledge local to each pre- and post-synaptic neuron pair; this means that our learning model presents many more neuromorphic characteristics than, for example, previous works where memristors are used as mere hardware implementations for artificial neural network weights (for example, see (\cite{10.1038/s42256-018-0001-4}).

Using memristive devices as weights can enable efficient computing - which is also one of the cornerstones of the neuromorphic approach - but not pairing them with a neuromorphic learning algorithm and a spiking neural network imposes quite stringent requirements on the physical characteristics of the device.
For example, a memristor simply used as a physical implementation of a network weight works best if it exhibits reliable, linear, and symmetric conductance response in order to approximate the idealised weight it stands in for (\cite{10.1038/s41467-018-04484-2}).
Present-day machine learning models suffer from “brittleness” (i.e., small changes in input can give rise to large errors) and do not seem well positioned in being able to match human adaptability across a wide variety of tasks.
It is becoming accepted that stochasticity is itself a computational resource (\cite{10.1038/s41563-019-0291-x}) - even though it is still not formally proven as such - and the fact that our model seems to perform best with a moderate amount of noise could be a result of this effect.
It had already been proposed that networks using memristors as synapses could match the learning performance of traditional artificial neural networks (\cite{10.1109/ted.2015.2439635}) and our results are in line with this conjecture.

\section*{Additional Requirements}

\subsection*{Contribution to the Field Statement}
Approaches based on artificial neural networks have found great success in performing a variety of tasks, but these are typically run on conventional (Von Neumann) computational architectures.  
Moving to a computational framework in which co-localisation of memory and computation is realised by brain-inspired (neuromorphic) hardware is a promising way to address the main limitations of current methods.
In our manuscript we simulated this approach by utilising novel interfacial memristive devices - whose behaviour resembles that of biological synapses - as weights in a neural network.
This network learned to represent non-linear functions through a training process based on an original supervised learning algorithm.
Using this class of memristive devices as the synaptic weight element in a spiking neural network yields, to our knowledge, one of the first models of this kind capable of learning to be a universal function approximator; their interesting dynamics and characteristics also strongly suggest the suitability of our memristors for usage in future neuromorphic computing platforms.
Our results are not limited to this class of memristors, but generalise to any whose behaviour follows a power law. 
We believe our findings and analyses also offer insight into building and benchmarking the algorithms necessary to benefit from these new materials.

\subsection*{Conflict of Interest Statement}
The authors declare that the research was conducted in the absence of any commercial or financial relationships that could be construed as a potential conflict of interest.

\section*{Author Contributions}
T.F.T., A.S.G., J.P.B., T.B. and N.A.T. developed the ideas and wrote the paper. 
A.S.G. conducted experimental work. 
T.F.T., A.S.G. and N.A.T. derived mathematical results. 
T.F.T. ran the simulations.

\subsection*{Acknowledgments}
A.S.G. and T.B. would like to thank all members of the Spintronics of Functional Materials group at the University of Groningen, in particular Arijit Das for help with device fabrication. Device fabrication was realised using NanoLab NL facilities. A.S.G. and T.B. acknowledge technical support from J. G. Holstein, H. H. de Vries, T. Schouten, and H. Adema. 
A.S.G. and T.F.T. are supported by the the CogniGron Centre, University of Groningen.

\section*{Data Availability Statement}
The experimental datasets generated during the current study are available from the corresponding authors on request. 
\section*{Code Availability}
All code used in this study is publicly available on GitHub at \url{https://github.com/Tioz90/Learning-to-approximate-functions-using-niobium-doped-strontium-titanate-memristors}

\section*{Abbreviations}
\begin{description}
\item[LIF: ]Leaky integrate-and-fire (neurons)
\item[mPES: ]memristor prescribed error sensitivity
\item[MSE: ]Mean squared error
\item[Nb: ]Niobium
\item[Nb:STO: ]Nb-doped SrTiO$_3$
\item[NEF: ]Neural Engineering Framework
\item[SNN: ]Spiking neural network
\item[SrTiO$\boldsymbol{_3}$: ]Strontium Titanate
\item[$\boldsymbol{\rho}$: ]Spearman correlation coefficient
\end{description}


\printbibliography

\clearpage
\renewcommand{\figurename}{Figure S}
\renewcommand\thefigure{\arabic{figure}}
\setcounter{figure}{0} 

\section*{Supplementary Data}


\begin{figure}[htbp]
\begin{center}
\includegraphics[width=\textwidth]{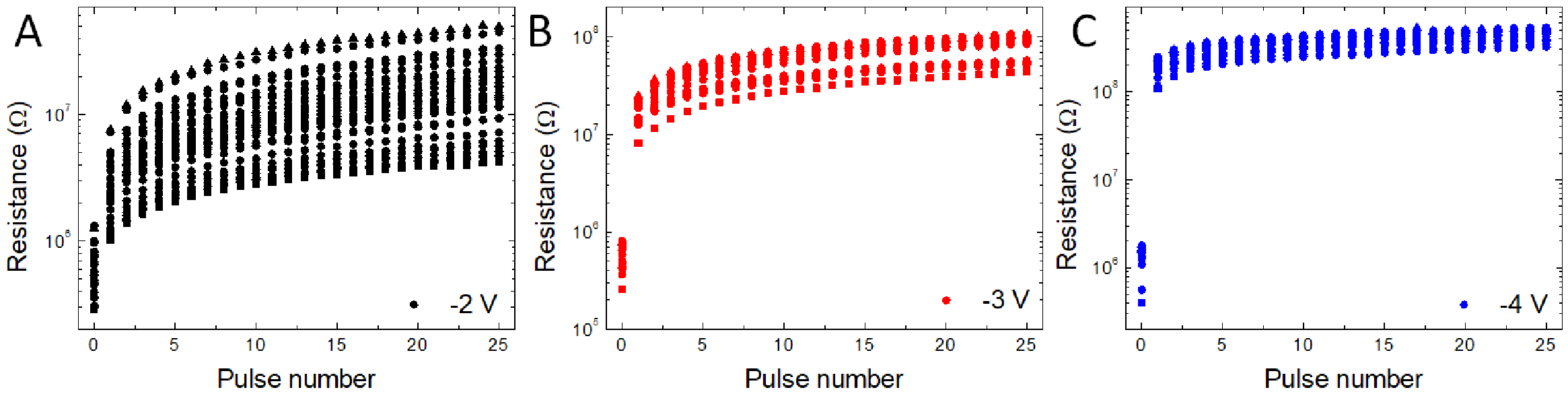}
\end{center}
\caption{Device response to RESET pulses of \textbf{(A)} -2 V, \textbf{(B)} -3 V and \textbf{(C)} -4 V (also shown in main text). Variation in the initial state is reflected by differences in the resistance at pulse number 0. In all cases, the first pulse induces the largest change and the effect of subsequent pulses is diminished.}
\label{fig:1}
\end{figure}

\begin{figure}[htbp]
\begin{center}
\includegraphics[width=\textwidth]{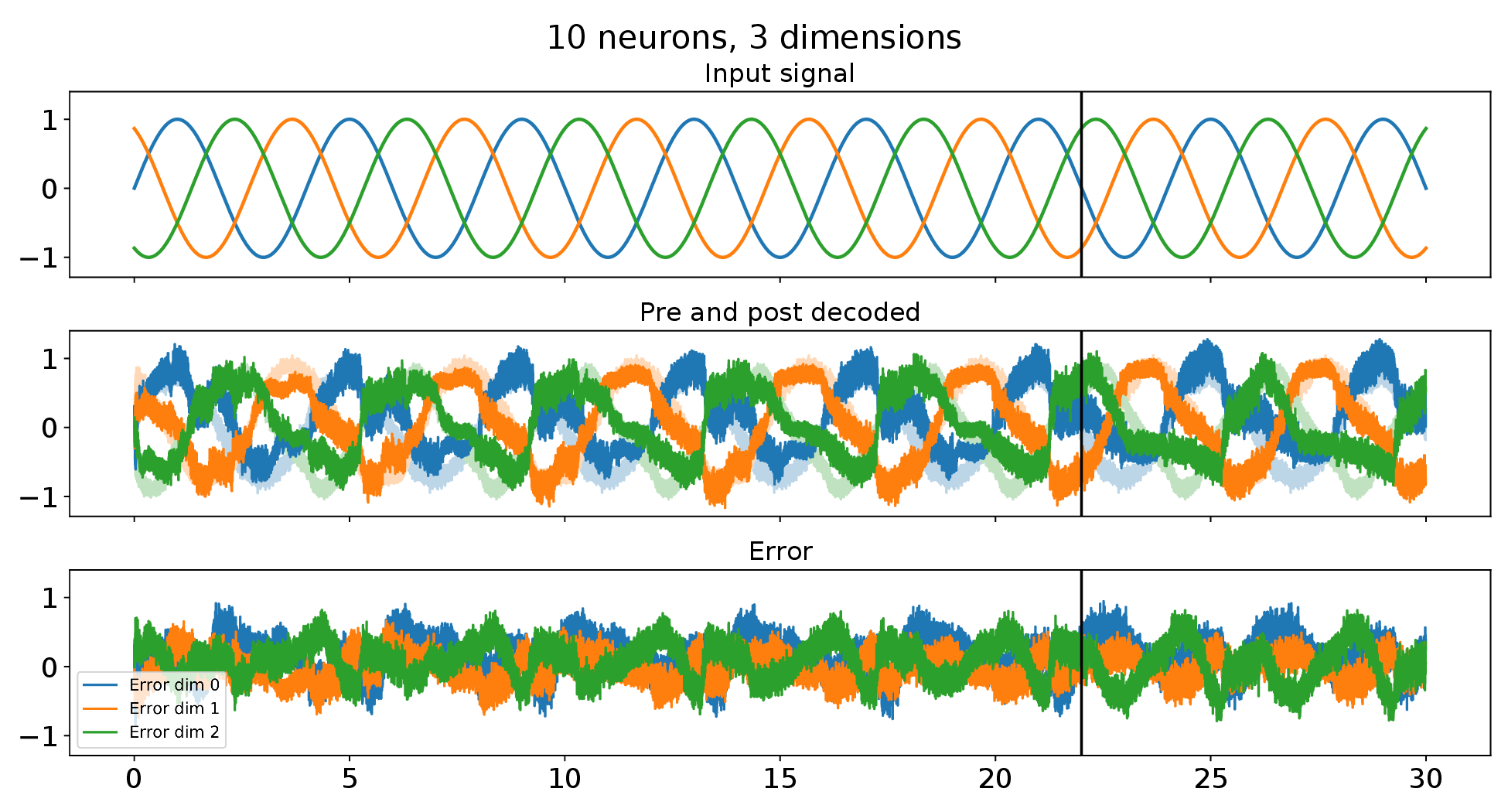}
\end{center}
\caption{
Example simulation during which a network composed of 10 pre- and post-synaptic neurons learns the identity function from a 3-dimensional sine wave.
After 22 seconds learning is switched off (vertical black bar) and the performance is tested on the same sine wave.
The top panel shows the input signal afferent to the pre-synaptic neuronal ensemble.
The middle panel shows the decoded output from the pre=synaptic ensemble (faded colours), and from the post-synaptic ensemble (bold colours).
The bottom panel shows the decoded output from the ensemble calculating the global error $\vec{E}$, which is used to drive learning.
}
\label{fig:2}
\end{figure}

\begin{figure}[htbp]
\begin{center}
\includegraphics[width=\textwidth]{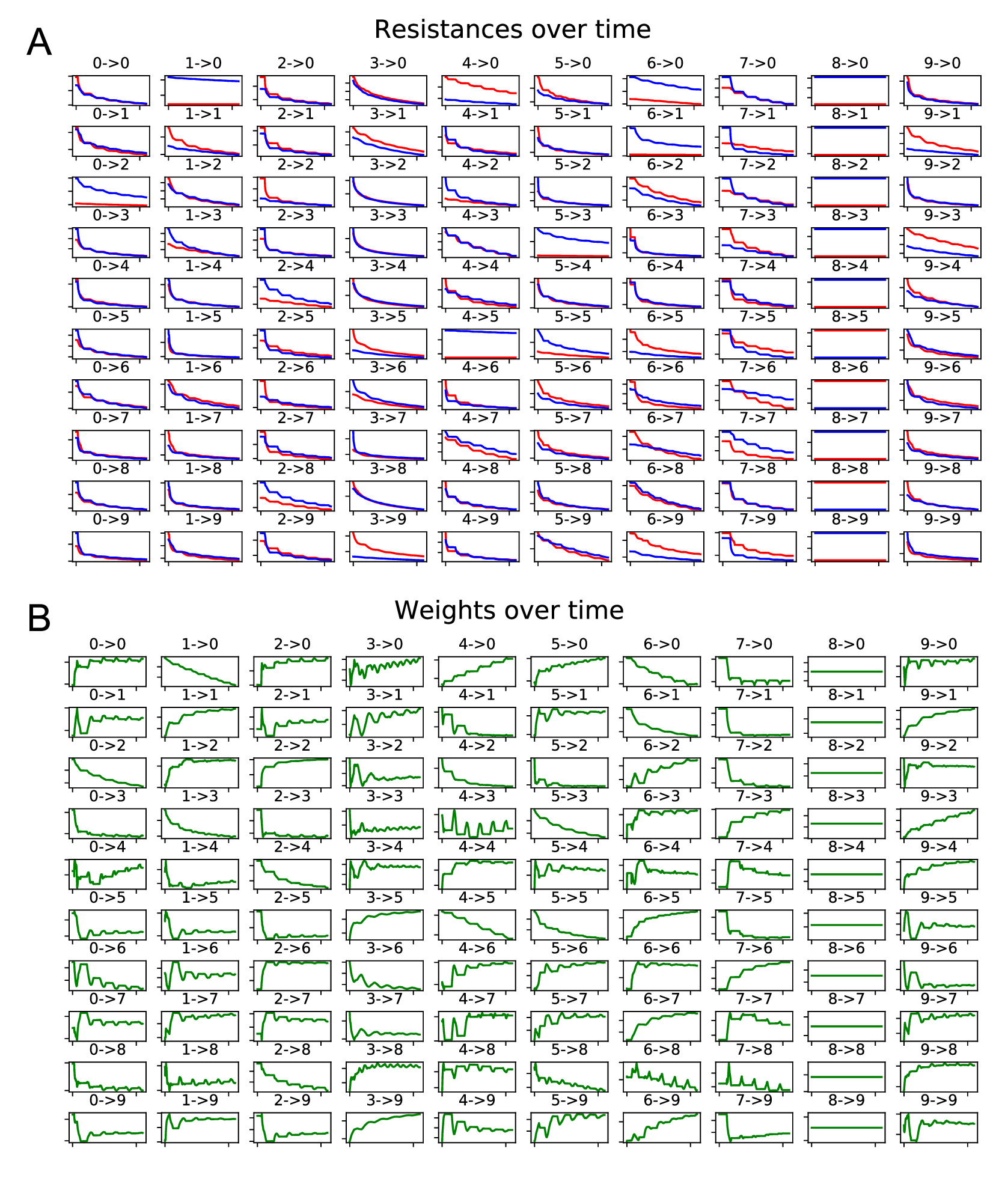}
\end{center}
\caption{
\textbf{(A)} Evolution of memristor resistances during the training phase for the network simulated in Fig. \ref{fig:2}.
The plots are organised such that the synapses efferent from pre-synaptic neuron $i$ are in column $i$, and the synapses afferent to post-synaptic neuron $j$ are in row $j$.
Each plot shows the positive (in red) $M_{ij}^+$ and negative (in blue) $M_{ij}^+$ memristor for the synapse connecting pre-synaptic neuron $i$ to post-synaptic neuron $j$.
\textbf{(B)} Evolution of the network weights $W_{ij}$ for each synapse connecting pre-synaptic neuron $i$ and post-synaptic neuron $j$. 
Each each weight is given by $W_{ij} = \zeta(M_{ij}^+ - M_{ij}^-)$, with $\zeta$ a transformation of the memristor resistances to network weights.
}
\label{fig:3}
\end{figure}



\begin{figure}[htbp]
\begin{center}
\includegraphics[width=\textwidth]{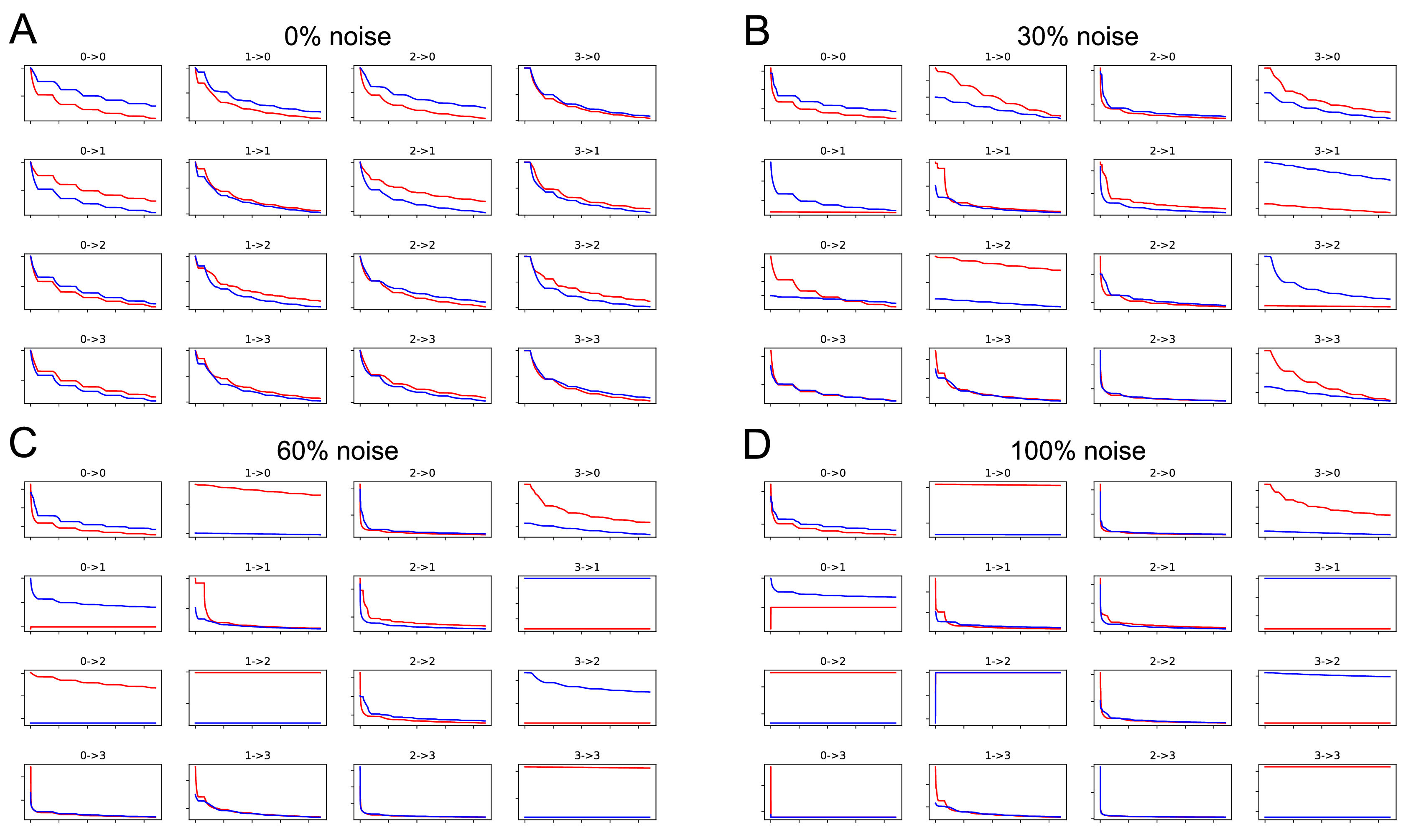}
\end{center}
\caption{
Evolution of memristor resistances during training for varying levels of coefficient of variation (expressed as percentage of noise) on the parameters $R_0$, $R_1$, $c$ in the power-law governing the resistance update, and initial resistances.
The variation was introduced via Gaussian random sampling.
The network for this example was run with 4 pre- and post-synaptic neurons.
\textbf{(A)} No noise added to parameters so each memristor in the network has the same update behaviour.  
All memristors start from the same HRS of $10^8 \, \Omega$.
\textbf{(B)} $30\%$ of coefficient of variation added to all parameters.
The memristors start from an initial resistance $[10^8 \, \Omega \pm 30\%]$ and vary in their response to learning SET pulses.
\textbf{(C)} $60\%$ of coefficient of variation added to all parameters.
Some memristors don't respond to pulses because their initial resistance was already equal to the minimum $R_0$, or to the maximum $R_1$.
\textbf{(D)} $100\%$ of coefficient of variation added to all parameters.
Many memristors don't respond, or stop responding because their resistance quickly saturates.
Some memristors see their resistance increase in response to SET pulses.
}
\label{fig:4}
\end{figure}

\end{document}